\documentclass[acmsmall,screen,]{acmart}

\usepackage{algorithm}
\usepackage{algorithmic}
\usepackage{graphicx}
\usepackage{subcaption}
\usepackage{textcomp}
\usepackage{multirow}
\usepackage[normalem]{ulem}
\usepackage{xcolor}
\usepackage{booktabs}
\usepackage{tabularx}
\usepackage{colortbl}
\usepackage{array}
\usepackage{enumitem}
\usepackage{xspace}
\usepackage{amsmath}
\usepackage{makecell}
\usepackage{tcolorbox}
\usepackage{array} 
\usepackage{adjustbox}
\usepackage{pifont}
\newcommand{\xmark}{\ding{55}} 

\newcolumntype{C}{>{\centering\arraybackslash}X}

\usepackage{tcolorbox}
\tcbuselibrary{breakable}
\usepackage{listings}
\usepackage[T1]{fontenc}   

\newtcolorbox{rqbox}{
    colback=gray!15,
    colframe=black,
    boxrule=0.5pt,
    arc=0pt,  
    left=8pt,
    right=8pt,
    top=5pt,
    bottom=5pt
}

\usepackage{framed}
\usepackage{xcolor}

\definecolor{shadecolor}{gray}{0.95}

\newtcolorbox{promptbox}[1][]{
    colback=gray!10!white,
    colframe=gray!75!black,
    title={#1},
    fonttitle=\normalfont\bfseries,
    breakable,
    left=5pt,
    right=5pt,
    top=3pt,
    bottom=3pt,
    before upper={\parindent0pt\footnotesize},  
}

\renewcommand\footnotetextcopyrightpermission[1]{}

\newenvironment{change}{\color{black}}{\color{black}}

\newcommand{\changetable}[1]{{\color{black}#1}}

\begin{document}

\title{VulInstruct: Teaching LLMs Root-Cause Reasoning for Vulnerability Detection via Security Specifications}


\author{Hao Zhu}
\affiliation{
  \institution{Peking University}
  \city{Beijing}
  \country{China}
}
\email{zhuhao@stu.pku.edu.cn}

\author{Jia Li}
\affiliation{
  \institution{Tsinghua University}
  \city{Beijing}
  \country{China}
}
\email{jia_li@mail.tsinghua.edu.cn}

\author{Cuiyun Gao}
\affiliation{
  \institution{Harbin Institute of Technology}
  \city{Harbin}
  \country{China}
}
\email{gaocuiyun@hit.edu.cn}

\author{Jiaru Qian}
\affiliation{
  \institution{Peking University}
  \city{Beijing}
  \country{China}
}
\email{qianjiaru77@gmail.com}

\author{Yihong Dong}
\affiliation{
  \institution{Peking University}
  \city{Beijing}
  \country{China}
}
\email{dongyh@stu.pku.edu.cn}

\author{Huanyu Liu}
\affiliation{
  \institution{Peking University}
  \city{Beijing}
  \country{China}
}
\email{huanyuliu@stu.pku.edu.cn}

\author{Lecheng Wang}
\affiliation{
  \institution{Peking University}
  \city{Beijing}
  \country{China}
}
\email{wanglecheng@stu.pku.edu.cn}

\author{Ziliang Wang}
\affiliation{
  \institution{Peking University}
  \city{Beijing}
  \country{China}
}
\email{wangziliang@pku.edu.cn}

\author{Xiaolong Hu}
\affiliation{
  \institution{New H3C Technologies Co., Ltd}
  \city{Hangzhou}
  \country{China}
}
\email{xlhu@h3c.com}

\author{Ge Li}
\affiliation{
  \institution{Peking University}
  \city{Beijing}
  \country{China}
}
\email{lige@pku.edu.cn}
\authornote{Corresponding author.}


\begin{abstract}

Large language models (LLMs) have achieved remarkable progress in code understanding and analysis tasks. However, state-of-the-art LLMs demonstrate limited performance in vulnerability detection tasks, and even state-of-the-art models struggle to distinguish vulnerable code from patched code. We argue that a key reason for this limitation is that LLMs lack an understanding of \textbf{security specifications}---the expectations defined by developers and security teams about how code should behave to remain safe. When the actual behavior of the code differs from these expectations and introduces a security risk, it becomes a potential vulnerability. However, such knowledge is rarely explicit in training data, leaving models unable to reason about the root causes of security flaws. 
To address this challenge, We propose \textbf{VulInstruct}, a specification-guided approach that systematically extracts reusable security specifications from historical vulnerabilities to instruct the detection of new ones. 
Specifically, VulInstruct designs two automatic pipelines to construct a \textbf{specification knowledge base} from complementary perspectives: 
(i) \textbf{General specifications}, extracted from high-quality patches across diverse projects, capturing fundamental safe behaviors accumulated across the open-source ecosystem; and (ii) \textbf{Domain-specific specifications}, context-dependent expectations repeatedly violated in particular repositories or domains that are relevant to the target code under analysis. 
Before analyzing new code, VulInstruct leverages this specification knowledge base to retrieve relevant past cases and their associated specifications, enabling LLMs to reason about expected safe behaviors rather than relying solely on surface patterns.
We evaluate VulInstruct under strict evaluation criteria requiring both correct predictions and valid reasoning. On the PrimeVul dataset, VulInstruct achieves 45.0\% F1-score (32.7\% improvement) and 37.7\% recall (50.8\% improvement) compared to the strongest baselines, while uniquely detecting 24.3\% of all identified vulnerabilities---2.4$\times$ more than any baseline. In pair-wise evaluation distinguishing vulnerable from patched code, VulInstruct also achieves a 32.3\% relative improvement over the best baseline. Beyond benchmarks, VulInstruct discovered a previously unknown high-severity vulnerability in production code (later assigned CVE-2025-56538) by recognizing violations of extracted specifications, demonstrating its practical value for real-world vulnerability discovery. All code and supplementary materials are available at https://github.com/zhuhaopku/VulInstruct-temp.

\end{abstract}




\maketitle

\section{Introduction}\label{sec:introduction}
Software vulnerability detection plays a critical role in software security, aiming to identify potential security flaws in source code that could be exploited by malicious attackers. With the rapid development of large language models (LLMs) and their success in code understanding and analysis tasks  \cite{li2025aixcoder7bv2trainingllmsfully,dong2025surveycodegenerationllmbased, jiang2024seedcustomizelargelanguage, zhang2023toolcoderteachcodegeneration, wang2023chatcoderchatbasedrefinerequirement}, recent studies have explored using LLMs for automated vulnerability detection \cite{nong2024chain,Tamberg_2025,ullah2024llmsreliablyidentifyreason,widyasari2024chatgptenhancingsoftwarequality, zhang2024promptenhancedsoftwarevulnerabilitydetection,zhou2024largelanguagemodelvulnerability, lekssays2025llmxcpg}. While these LLM-based approaches have achieved notable improvements over traditional methods, their performance on vulnerability detection benchmarks remains unsatisfactory. Recent evaluation \cite{ding2024vulnerability} shows that even state-of-the-art LLMs achieve less than 12\% accuracy in distinguishing vulnerable from patched code, highlighting a fundamental limitation in understanding the root causes of vulnerabilities.

We argue that one key reason behind the limited performance of LLMs is that models lack an understanding of the \textbf{security specifications} in code. A security specification is the expectations defined by developers and security teams about how the code should behave in order to remain safe. 
When the actual behavior of the code differs from this expectation and introduces a security risk, it becomes a potential vulnerability. The following example illustrates this fundamental challenge:

\begin{tcolorbox}[colback=gray!5!white, colframe=gray!75!black, title=Example: TLS Certificate Validation]
Consider a case where a TLS library fails to verify the hostname of a server during certificate validation. To a developer who is unaware of the relevant specification, this code might appear correct: the function successfully completes a TLS handshake and retrieves the server certificate. However, the expected code behavior requires that the hostname in the certificate must be checked against the intended server identity. Only when this expectation is made explicit does it become clear that the missing check introduces a serious security flaw, since an attacker could impersonate any server with a valid certificate.
\end{tcolorbox}

However, such specifications are rarely documented explicitly in code or public resources. In practice, they are typically discussed and agreed upon during design or review phases, and prior work \cite{wan2024bridging} has noted that this knowledge is often shared through internal training or reviews. As a result, LLMs trained primarily on code rarely encounter these implicit expectations and thus struggle to reason about the root causes of vulnerabilities.

Our key motivation is that historical vulnerabilities implicitly encode the security specifications defined by developers and security teams (we provide a detailed discussion and illustrative cases in Section~\ref{sec:motivation}). Based on this insight, we propose \textbf{VulInstruct}, a specification-guided approach for vulnerability detection. The central idea of VulInstruct is to extract \textbf{reusable security specifications from historical vulnerabilities} and use them to instruct the detection of new ones. Before analyzing a target code snippet, VulInstruct retrieves similar historical vulnerability cases and provides their associated specifications as explicit security knowledge. Under the support of such specifications, LLMs can go beyond common or surface defect patterns and reason about whether the code violates an expected safe behavior.

A key technical challenge is how to extract security specifications in a systematic way. To address this,  VulInstruct designs two automatic pipelines to construct a \textbf{specification knowledge base} from complementary perspectives: 
\textbf{(i) General security specifications.} Extracted from high-quality patch datasets across diverse projects, a focus that has long been central to the vulnerability detection community, these specifications capture fundamental expected behaviors representing security expertise accumulated across the entire open-source ecosystem.
We then extract them from high-quality patch datasets by comparing vulnerable code with its fixed version and restating the underlying expected behaviors as explicit, reusable specifications. 
To move beyond prior works, we further capture surrounding information such as callee functions, type declarations, imported modules, and global variables alongside the vulnerable and patched code, enabling more precise abstraction of the developer’s intended safe behavior. 
\textbf{(ii) Domain-specific security specifications}: 
Dynamically extracted from our comprehensive CVE database, these specifications are derived from frequently exploited vulnerabilities within the same repository or related projects in the same domain of the target detected code. 
By studying how attackers repeatedly exploit similar weaknesses, we identify context-specific expectations whose violations enable attacks. These specifications capture the attacker's perspective, revealing which expected behaviors matter most in practice for specific codebases.
Together, the two types of specifications are complementary: general specifications provide broad coverage of fundamental secure behaviors, while domain-specific specifications capture context-aware expectations that matter most in practice.

We evaluate VulInstruct under the stricter evaluation framework \cite{li2025everything}, which requires not only correct predictions but also valid reasoning. 
On the widely adopted PrimeVul dataset \cite{ding2024vulnerability}, VulInstruct achieves 45.0\% F1-score, representing a 32.7\% relative improvement over the strongest baseline. VulInstruct's 37.7\% recall substantially outperforms all baselines, with 24.3\% of its detections being unique vulnerabilities that other methods miss entirely. In pair-wise evaluation, where models must correctly distinguish vulnerable code from its patched code, VulInstruct achieves 17.2\% P-C accuracy, improving by 32.3\% over the best baseline. Our analysis further reveals that effective vulnerability detection requires not just access to security knowledge but careful selection of the most relevant specifications and cases. 
Beyond controlled evaluations, we demonstrate VulInstruct's real-world utility through a case study where our specification-guided approach successfully identified a previously unknown high-severity vulnerability. The vulnerability, which violated the same security specification as a known CVE, was confirmed and subsequently patched by developers, highlighting VulInstruct's potential for practical vulnerability discovery.

Our contributions are summarized as follows:
\begin{itemize}
    \item \begin{change}We propose a novel perspective on LLM-based vulnerability detection by systematically mining reusable \emph{security specifications} from historical vulnerabilities. Unlike existing retrieval-based approaches that provide concrete vulnerability cases as reference, we extract natural-language specifications that capture implicit developer knowledge about expected safe behaviors, enabling direct use as LLM reasoning context.\end{change}
    
    \item We propose VulInstruct, a novel specification-guided approach that extracts implicit knowledge from historical vulnerabilities through a dual-layer retrieval design and leverages it to instruct LLMs in vulnerability detection.
    \item Under more strict evaluation, VulInstruct achieves state-of-the-art vulnerability detection performance, improving F1-score by 32.7\% and recall by 50.8\% over the strongest baselines. 
\end{itemize}

\vspace{-0.3cm}

\section{Background and Related Work}\label{sec:background}

The limitations of LLM-based vulnerability detection were most clearly articulated by the PrimeVul benchmark\cite{ding2024vulnerability}. The authors concluded that even state-of-the-art LLMs achieve less than 12\% accuracy in distinguishing vulnerable code from patched code. Their discussion explicitly highlighted three challenges for future research: (i) the need for richer program context, (ii) the importance of teaching models to reason about vulnerability detection beyond binary labels, and (iii) the lack of security awareness in current models. Subsequent work has largely followed these directions.

\subsection{Code Context and Reasoning in Vulnerability Detection}
\label{relate_work_eval}


\textbf{Leveraging richer program context.} Recent studies \cite{wang2024reposvul, wen2024vuleval} have identified code context as a critical factor in vulnerability detection. Risse et al. \cite{risse2025top} reported that most of the functions in widely used benchmarks cannot be reliably classified without knowing of how they are invoked and used in the program. They enumerated several types of important contextual information, such as external functions, type declarations, and global variables. More recent approaches have explored various strategies to leverage richer program context. Yildiz et al. \cite{yildiz2025jitvul} uses ReAct agents for recursive callee retrieval, Yang et al. \cite{yang2025context} abstracts primitive API calls at multiple granularity levels. Wang et al. \cite{wang2025savant} introduced Savant, which uses LLM semantic understanding to detect vulnerable API usages in Java dependencies through reflection-based iterative context expansion. Shemetova et al. \cite{shemetova2025lamed} leveraged LLMs to automatically generate function annotations for memory allocation/deallocation patterns in code context to integrate with static analyzers.

\textbf{Teaching models to reason about vulnerability detection.}
In response to PrimeVul’s second challenge, a growing line of work has  focused on enhancing models' reasoning capabilities in vulnerability detection.
One direction is to use prompting strategies and in-context learning. 
These approaches \cite{lu2024grace, sun2024llm4vuln, yang2025dlap,li2025llmsjustlooksimple, nong2024chain} typically leverage program analysis techniques such as control flow graphs (CFG), data flow analysis, and program dependency graphs (PDG) to construct prompts that stimulate models' exploration capabilities. Beyond prompting, training-based methods often decompose vulnerability detection into subtasks to improve reasoning quality. Du et al. \cite{du2024generalization} proposed VulLLM, which uses multi-task learning across detection, localization, and interpretation. Weyssow et al. \cite{weyssow2025r2vul} introduced R2Vul, combining reinforcement learning with structured reasoning distillation to train small LLMs for vulnerability detection with explanations. 
Wen et al. \cite{wen-etal-2025-boosting} proposed ReVD, which applies triplet supervised fine-tuning with curriculum preference optimization to capture vulnerability patterns.
A third line of work leverages multi-round, multi-agent interactions to simulate structured reasoning. Hu et al. \cite{hu2023large} proposed GPTLens with auditor agents generating vulnerabilities and critic agents evaluating them adversarially. Widyasari et al. \cite{widyasari2025let} introduced VulTrial, a courtroom-inspired multi-agent framework featuring four role-specific agents, achieving 102.39\% improvement over single-agent baselines with GPT-4o.

\textbf{Despite these advances, most existing studies still evaluate models using binary labels}: whether code is vulnerable or not. Such evaluation is unsatisfactory for LLMs: a model may predict the correct label but rely on spurious correlations rather than genuine understanding of root causes. Consequently, researchers~\cite{kaniewski2025systematic, li2025irisllmassistedstaticanalysis} have increasingly advocated for reasoning-aware evaluation. Li et al.~\cite{li2025everything} proposed the CORRECT framework, which requires both the binary classification label and the reasoning process to be correct. By leveraging sufficient code context, they demonstrated that stricter evaluation can reveal whether models truly understand vulnerability mechanisms. As LLMs’ reasoning capabilities continue to improve, more works are shifting toward such stricter metrics. This trend motivates our adoption of CORRECT evaluation in this paper.



\subsection{Security Awareness in LLM-based Detection}

\textbf{Augmenting Security Knowledge.} PrimeVul also emphasized that current code LMs lack what the authors termed “security awareness.” 
In their discussion, this referred to the observation that models often decide based on textual similarity in training dataset rather than considering the underlying causes of vulnerabilities or fixes of the vulnerabilities. 
They suggested that researchers should explore ways to teach LMs security knowledge, inspired by how human developers are trained in secure coding. 

In practice, however, such knowledge rarely exists explicitly. 
Security expertise in industry is often implicit: experts accumulate best practices through hands-on experience, and then transfer this knowledge to developers through periodic training sessions or internal guidelines~\cite{wan2024bridging}. Yet no comprehensive dataset currently exists that systematically captures and encodes such security knowledge for AI models. 
Several efforts have begun to make such implicit security knowledge more explicit. BugScope \cite{guo2025bugscope} leverages LLMs to summarize curated vulnerability cases, thereby expanding code patterns associated with common vulnerability types. 
In the formal analysis community, APHP \cite{lin2023detecting} and Seal \cite{chen2025seal} recover API usage constraints from security patches, identifying rules such as “\texttt{release()} must follow \texttt{acquire()}.” These studies demonstrate that structured rules can indeed be distilled from developer fixes, turning implicit practices into explicit specifications. 
\begin{change}
A key distinction of our work is the choice of natural language as the specification formalism. Prior formal approaches and SAST rules express constraints as structural patterns over program dependence graphs or API call sequences. While precise, such representations require specialized program analysis expertise to construct and are tightly coupled to specific code structures, limiting their transferability across projects. In contrast, our specifications describe expected safe behaviors in natural language. We argue that this formalism offers two key advantages for LLM-based detection: (i) \textbf{Easier acquisition}---specifications are automatically extracted by prompting LLMs on patches, without requiring specialized program analysis expertise or infrastructure; and (ii) \textbf{Higher-level abstraction}---the natural-language form describes expected safe behaviors rather than code-structural patterns, reducing dependence on concrete program structures and improving cross-project generalization (as demonstrated in Section~\ref{sec:learning-from-patches}, where a specification extracted from libgadu generalizes to e2guardian).
Furthermore, natural language is the native modality that LLMs reason in, allowing direct integration of specifications into the detection context without format translation.

\end{change}

In contrast, most existing methods in vulnerability detection, including Vul-RAG~\cite{du2024vul} and the training-based approaches discussed above, primarily rely on exposing models to more vulnerability examples through retrieval or fine-tuning. While this can improve recognition of familiar patterns, it does not provide models with the underlying expectations about safe behavior that truly determine whether code is vulnerable. Consequently, models remain limited to pattern matching rather than reasoning with explicit security knowledge. A systematic way to extract and apply such structured security knowledge has been lacking, motivating the design of our approach.

\section{Motivation}\label{sec:motivation}


\label{sec:security-specs}

As introduced in Section~\ref{sec:introduction}, vulnerabilities fundamentally arise when an implementation violates a security specification. We define a security specification as the expectation about safe program behavior established by developers and security experts. These expectations are rarely documented explicitly, but can be reconstructed from historical evidence. We identify two complementary perspectives for recovering such expectations: \textbf{(i) General specifications:} By analyzing how vulnerable code differs from its fixed version across high-quality patches, we can explicitly restate the expected safe behavior that was violated. \textbf{(ii) Domain-specific specifications:} By studying vulnerabilities that share similar exploitation mechanisms within the same domain, we can recover the underlying expectations that were repeatedly violated.

A key \textit{insight} is that every vulnerability---no matter how local or project-specific---can be 
traced to the violation of at least one underlying security specification. 
For example, failing to reset session state after authentication errors violates an expectation 
of state consistency; using freed memory without nullification violates an expectation about 
pointer lifecycle; neglecting TLS hostname verification violates an expectation about protocol 
boundary enforcement. These expectations, which we formalize as \textit{security specifications}, 
serve as a unifying representation of implicit expert knowledge, making them both interpretable 
to humans and transferable across projects.

\subsection{Learning General Specifications: TLS Hostname Verification}\label{sec:learning-from-patches}
\begin{figure}[t]
    \centering
    \includegraphics[width=\linewidth]{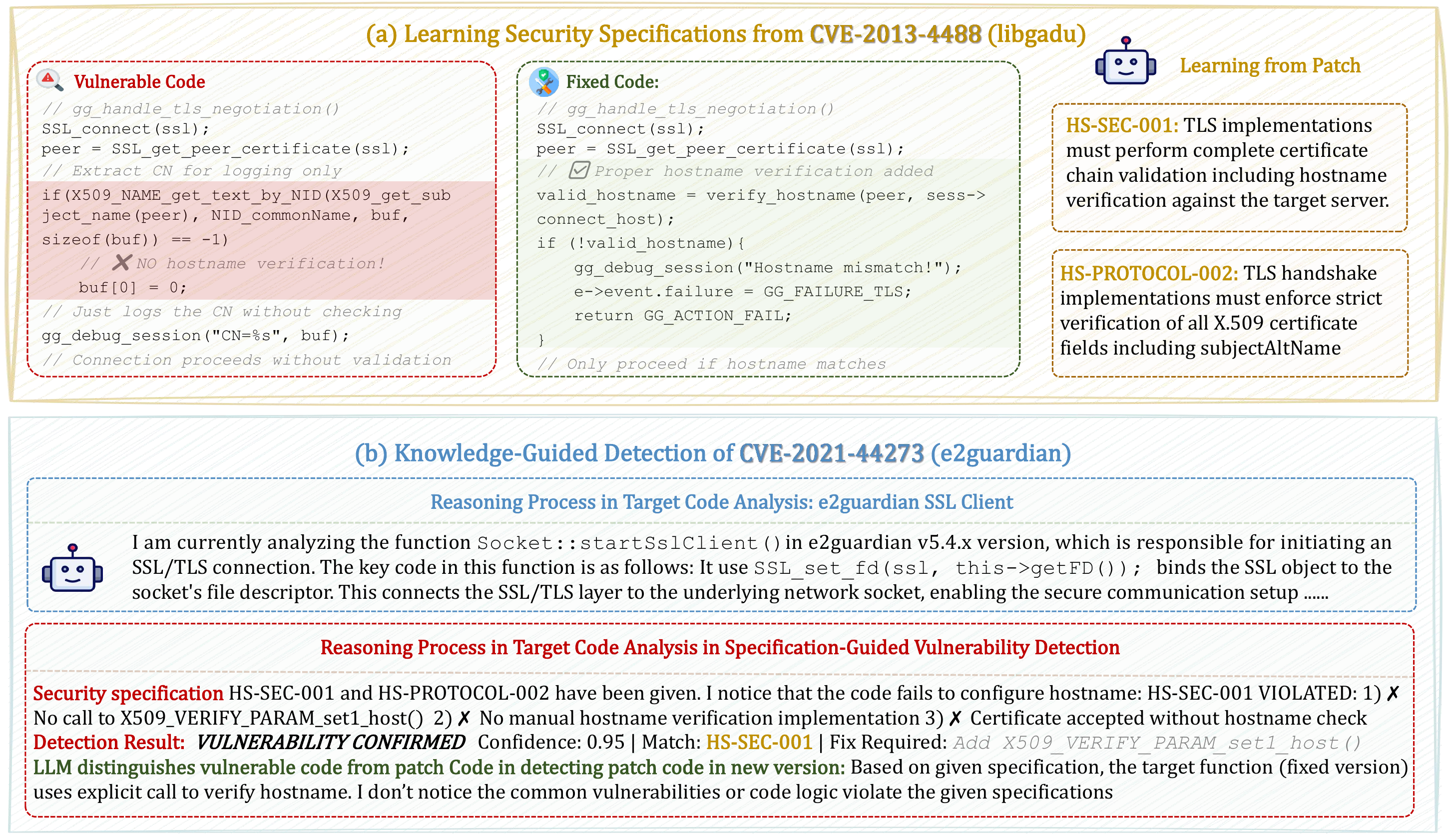}
    \caption{(a) CVE-2013-4488 vulnerability analysis: The vulnerable code 
    path in \texttt{libgadu} showing missing hostname verification, and VulInstruct's automated specification extraction from the patch. (b) VulInstruct using specifications from historical cases to detect new vulnerability}
    \label{fig:vulinstruct-specs}
    \vspace{-0.58cm}
\end{figure}

\textit{Case Study: TLS Hostname Verification.} The importance of hostname verification in TLS has been recognized for decades. RFC 2595 (1999) \cite{newman1999using}  already mandated that “the client MUST check its understanding of the server hostname against the server’s identity… to prevent man-in-the-middle attacks”, later consolidated in RFC 6125 (2011) \cite{saint2011representation}. Despite being standardized, such requirements rarely appear in project documentation and remain security experts’ implicit knowledge base. In practice, this specification is often overlooked. Figure~\ref{fig:vulinstruct-specs}(a) shows CVE-2013-4488 in the \texttt{libgadu} library, where the TLS negotiation handler retrieved and logged the server certificate but failed to verify it against the target hostname. This omission allowed attackers with any valid certificate—even self-signed—to impersonate arbitrary servers, completely undermining TLS security.

\textit{VulInstruct’s Automated Specification Learning.}
From the patch that fixed CVE-2013-4488 by adding hostname verification, VulInstruct automatically extracts the underlying security principles. Specifically, it inferred two structured specifications: \textbf{(i) HS-SEC-001}: TLS implementations must perform complete certificate chain validation including hostname verification.
\textbf{(ii) HS-PROTOCOL-002}: TLS handshakes must enforce strict verification of all X.509 certificate fields. This illustrates how VulInstruct turns implicit expert rules into explicit, LLM-readable specifications, bridging the gap between human reasoning and model inference.

\textit{Specification-guided Detection.}
Once extracted, these specifications become reusable knowledge. For example, when analyzing \texttt{e2guardian} (CVE-2021-44273), VulInstruct retrieved HS-SEC-001 and successfully detected a similar omission in the function \texttt{Socket::startSslClient()}, where no hostname verification was performed for OpenSSL (Figure~\ref{fig:vulinstruct-specs}b). The system flagged the violation and explained the attack impact, confirming the vulnerability with high confidence. This case demonstrates how specifications distilled from historical patches can generalize across projects, enabling specification-guided detection of previously unseen vulnerabilities.

\subsection{Learning Domain-specific Specifications: Image Parser Vulnerabilities}
\label{sec:learning-from-patterns}

\begin{figure}[t]
    \centering
    \includegraphics[width=0.95\linewidth]{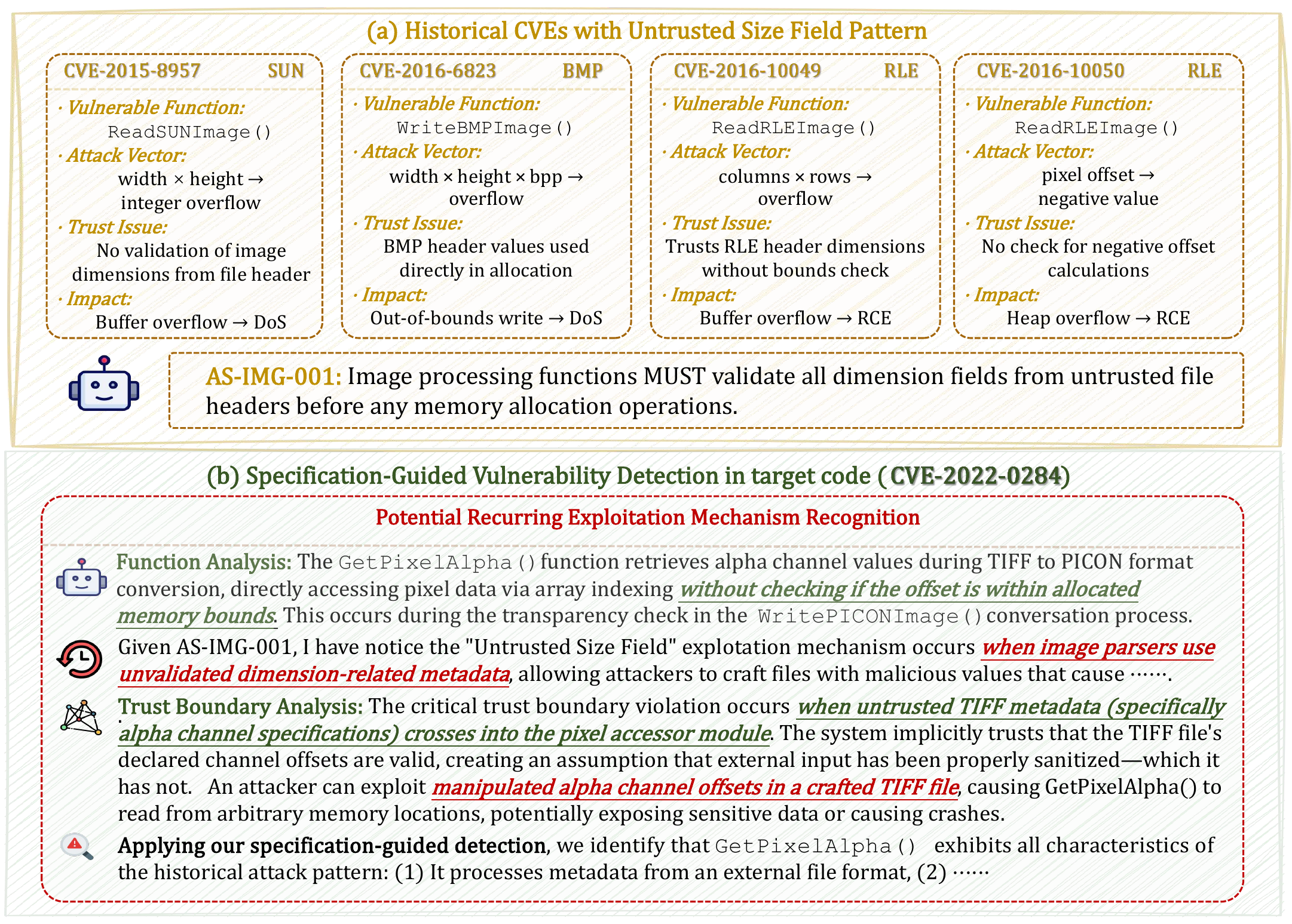}
    \caption{Domain-specific Recurring exploitation mechanism in ImageMagick: unvalidated length fields across historical and new vulnerabilities.}
    \label{fig:imagemagick-attack}
    \vspace{-0.5cm}
\end{figure}

Attackers rarely invent entirely new exploits; instead, they frequently 
\textit{reuse exploitation strategies} that have succeeded in the past. 
Google Project Zero reported that among 18 zero-day vulnerabilities disclosed in 2022, at least half were variants of previously patched vulnerabilities~\cite{googlezero2022}. This recurrence shows that vulnerabilities often stem from the same underlying weakness,  which attackers repeatedly target across projects in the same domain. 

Why can such exploits be reused? Because each recurring exploitation mechanism corresponds to the violation of the similar \textbf{security specification}---an implicit expectation about safe behavior that was never properly enforced. In this sense, attack patterns are not separate from security specifications, but evidence of which specifications matter most in practice. By abstracting recurring exploitation mechanisms into explicit specifications, we capture the security expectations needed to prevent the same class of attacks from reappearing and make them reusable knowledge for guiding vulnerability detection.

\textit{Case Study: Image Parsing in ImageMagick.}
Figure~\ref{fig:imagemagick-attack} illustrates how similar exploitations emerge in practice. Figure (a) summarizes a sequence of historical CVEs (2015–2016) in ImageMagick image parsers. Despite occurring in different formats (SUN, BMP, RLE), these vulnerabilities share a recurring exploitation mechanism: \textbf{untrusted size fields from image headers were directly used in allocation or offset calculations.} For example, CVE-2015-8957 trusted width × height values in the SUN image parser, leading to buffer overflow, while CVE-2016-6823 and CVE-2016-10049 in BMP and RLE coders suffered similar overflows due to unchecked dimension metadata. Across cases, the common thread is the same: failure to validate external size metadata before use.

\textit{Applying Domain-specific Specifications.}
Figure~\ref{fig:imagemagick-attack}b illustrates how VulInstruct leverages retrieval and abstraction during the analysis of CVE-2022-0284. When examining the function \texttt{GetPixelAlpha()} in the TIFF-to-PICON conversion pipeline, the system first retrieved historical CVEs with similar characteristics where untrusted dimension metadata flowed directly into memory access (as shown in Figure~\ref{fig:imagemagick-attack}a). VulInstruct then identified a consistent exploitation mechanism: “Untrusted Size Field” exploitation pattern and concluded AS-IMG-001 specification. Finally, VulInstruct applied this specification back to the target code, constructing a threat model that explained how manipulated alpha channel offsets in a crafted TIFF file could trigger out-of-bounds access.


\section{Methodology}\label{sec:methodology}

\begin{figure}[htbp]
    \centering
    \includegraphics[width=0.95\linewidth]{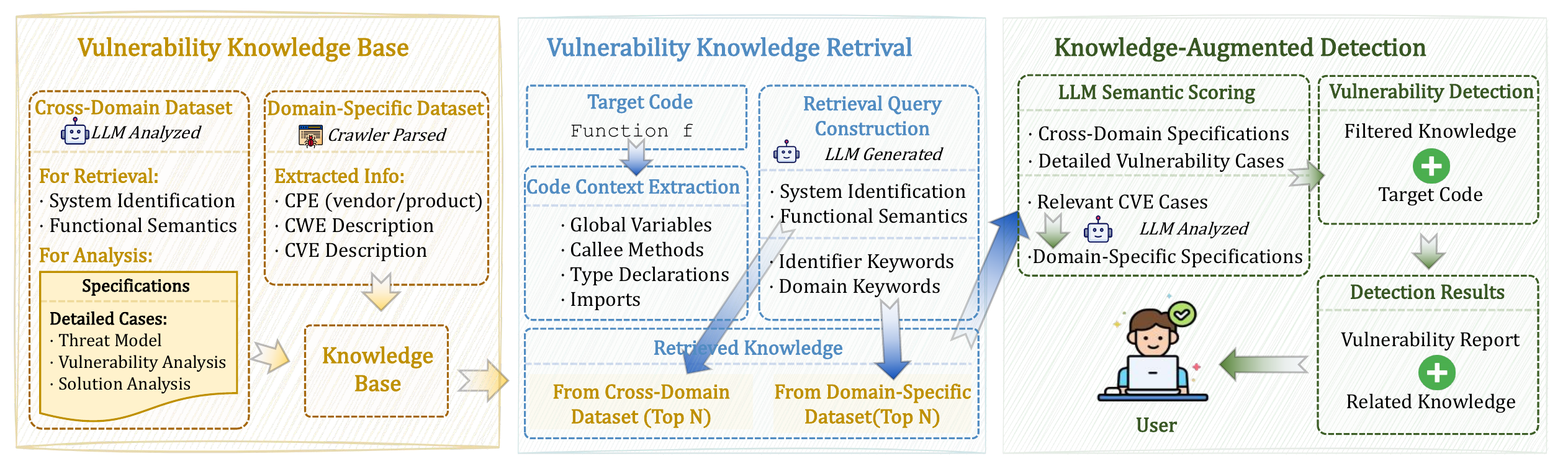}
    \caption{Overview of VulInstruct}
    \label{fig:architecture}
    \vspace{-0.5cm}
\end{figure}

Figure~\ref{fig:architecture} shows the workflow of \textbf{VulInstruct}, which combines \emph{offline} knowledge construction with \emph{online} specification-guided detection. Offline, VulInstruct designs two automatic pipelines to build a \textbf{Specification Knowledge Base (SKB)} that includes (i) a \emph{general specification knowledge base}, consisting of reusable specifications extracted from high-quality patches across diverse projects, and (ii) a large-scale \emph{domain evidence base}, collected from the CVE database, which provides relevant vulnerability cases for deriving domain-specific specifications. 
\emph{Online}, given a target code snippet, VulInstruct performs dual-path retrieval: spec-level retrieval over general specifications, and case-level retrieval over the domain evidence base. From the retrieved cases, it dynamically induces domain-specific specifications. Finally, VulInstruct performs specification-guided reasoning to detect vulnerabilities with explanations.
\begin{change}
The remainder of this section describes each phase: 
Section~\ref{subsec:general_specification_knowledge_base} and~\ref{subsec:domain_evidence_base} 
detail the offline construction of the specification knowledge base (Section~4.1), 
Section~4.2 presents the dual-path knowledge retrieval, 
and Section~\ref{sec:skb} describes the specification-guided detection pipeline.
\end{change}

\subsection{Specification Knowledge Base}

The SKB consists of two complementary parts. 
\textbf{general specifications} are pre-extracted, reusable rules of expected safe behavior, constructed offline from high-quality patches across diverse projects. In contrast, \textbf{domain-specific specifications} are not pre-built; instead, we maintain a domain evidence base collected from the CVE database, and during detection, VulInstruct dynamically abstracts domain-specific specifications from the most relevant retrieved cases. 
This design allows general specifications to provide broad, cross-project coverage, while domain-specific specifications capture context-aware expectations directly relevant with the target code.

\subsubsection{General Specification Knowledge Base}
\label{subsec:general_specification_knowledge_base}

The general specification knowledge base in VulInstruct contains two kinds of knowledge: (i) \emph{general specifications}, reusable rules that abstract the expected safe behavior, and 
(ii) \emph{detailed vulnerability cases}, structured analyses that ground each specification in concrete vulnerability–patch examples. Together, they capture both general, reusable specifications and the concrete detail from real-world codebases.

\noindent\textbf{General specifications.}  
A general security specification is designed as follows:
\begin{equation}
\label{eq:domain}
\text{HS-\texttt{[DOMAIN]}-\texttt{[ID]}}:\text{[Expected Safe Behavior]}
\end{equation}
where \begin{change}\texttt{HS} stands for Historical Specification\end{change}, \texttt{DOMAIN} indicates the security domain (e.g., MEM for memory safety, PROTOCOL for communication protocols, STATE for state consistency), and \texttt{ID} provides a unique index. The requirement is always expressed in an \emph{Expected Safe Behavior} form (``must do X'' rather than ``must not omit X''), ensuring both clarity and reusability. 

\noindent\textbf{Detailed vulnerability cases.}  
Each specification is linked with a structured case that grounds the abstract rule in concrete evidence from real code. Each case is documented in three forms:

\begin{itemize}[leftmargin=1.1em]
  \item \emph{Threat model.} A system-level view that clarifies trust boundaries, attack surfaces, and the vulnerability chain (e.g., how an initial flaw escalates into a critical CWE). 
  This highlights why the specification matters in the broader security context, ensuring that the rule is not only syntactic but rooted in realistic attack scenarios. 
  \item \emph{Vulnerability analysis.} Code analysis that traces the vulnerable execution path, identifies entry points and preconditions, pinpoints the flawed logic, and maps the violation back to the specification, which provides precise evidence of how the specification is broken in practice.

  \item \emph{Solution analysis.} Patch explanation that describes how the applied fix restores compliance with the violated specification, explicitly linking code changes to expected safe behaviors. This step confirms that the specification not only explains the cause of the vulnerability but also captures the principle underlying its fixing.

\end{itemize}

Together, these analyses complement the reusable specification itself: the specification captures the expected safe behavior, while the linked case provides concrete evidence of its violation and enforcement in real-world code.

\noindent\textbf{Full automatic pipeline.}  
We extract specifications from high-quality patch datasets. Each vulnerability instance is represented as
$D_i = \{c_i^{\text{vul}}, c_i^{\text{fix}}, m_i, d_i, w_i, \mathrm{Ctx}(f_i)\}$, 
where $c_i^{\text{vul}}$ and $c_i^{\text{fix}}$ are the vulnerable and fixed functions, $m_i$ is the commit message, $d_i$ the CVE description, $w_i$ the CWE type, and $\mathrm{Ctx}(f_i)$ the broader program context:
\begin{equation}
\label{eq:context}
\mathrm{Ctx}(f) = \{\mathrm{callees}(f, d), \ \mathrm{types}(f), \ \mathrm{imports}(f), \ \mathrm{globals}(f)\}.
\end{equation}
This representation ensures that each instance is grounded not only in the function body but also in its interaction with the surrounding system.  

Then, each general specification is inferred by prompting LLMs to perform reasoning on each vulnerability instance. We guide the LLM through a carefully designed one-shot prompt, \begin{change}which follows a fixed workflow: system understanding (identifying the software system, domain, and module), domain classification (categorizing the vulnerability into one specific security domain), and specification abstraction (reasoning about the root cause and the corresponding expected safe behavior). \end{change} Given the extracted general specifications, we further prompt the LLM to generate detailed vulnerability cases that link each specification to concrete evidence in the code.

\noindent\textbf{Retrieval keys.}  
Finally, to enable efficient matching with new code, we attach two retrieval keys to each specification:  
(i) a \emph{system identification key} describing the subsystem, module, and interactions of the function, and  
(ii) a \emph{functional semantics key} summarizing its intended behavior.  
Combined with the broader context in Equation~\ref{eq:context}, these keys allow VulInstruct to retrieve relevant specifications based on both program role and semantics, rather than relying on surface similarity.

\subsubsection{Domain Evidence Base}
\label{subsec:domain_evidence_base}

We construct a domain evidence base to capture recurring exploitation patterns in CVE cases. Unlike general specifications, which require high-quality patch datasets, domain-specific cases rely on breadth rather than depth. For this purpose, we collect all publicly available CVEs from the National Vulnerability Database (NVD)~\cite{nvd2024}, which is maintained by the U.S. National Institute of Standards and Technology (NIST) and provides comprehensive vulnerability metadata including severity scores, impact metrics, and reference information for each CVE entry. 
We design an automatic pipeline that crawls and normalizes CVE data from the National Vulnerability Database. In this work, we collect entries published between 2002 and 2024, excluding rejected ones. This process yields 173{,}070 valid cases and 15{,}391 rejected cases. For each CVE, the pipeline automatically normalizes key fields, including the vulnerability description, its associated CWE type, Common Platform Enumeration(CPE) identifiers (vendor and product names), reference links to advisories and patches, and the publication date for temporal filtering.

\subsection{Knowledge Retrieval}
\begin{change}
Given a target function $f$ to be analyzed, VulInstruct performs online retrieval 
over the pre-built specification knowledge base to identify the most relevant 
knowledge for guiding detection. The retrieval operates along two parallel paths, 
corresponding to the two components of the SKB.
\end{change}
The Vulnerability Knowledge Base (VKB) contains two complementary layers of knowledge: \emph{general security specifications} and  \emph{domain-specific security specifications}. 
Given a target function $f$ to be analyzed, the retrieval process aims to identify the most relevant knowledge items from the VKB that can guide subsequent detection.

\textbf{Code Context Extraction.}
We first implement an automatic tool that, given a target function $f$, extracts its surrounding program context as defined earlier in Equation~\ref{eq:context}. 

The extracted context is then incorporated into the retrieval query construction to improve semantic alignment with historical vulnerabilities in knowledge bases.

\textbf{Retrieval Query Construction for general specifications.}
To build effective retrieval queries, we prompt LLM to generate two forms of system-level descriptions for the target function:
\emph{system identification}, which specifies the subsystem, module, and component interactions where $f$ resides, and 
\emph{functional semantics}, which summarize the purpose and detailed behavior of $f$. We then use \textbf{Embedding-based Retrieval.} We concatenate the target code, the system identification and functional semantics of $f$, and encode them into dense embeddings. Using embedding similarity search, we retrieve the most relevant \emph{top-$k$ cases}: $\{K^{(\text{spec})}_{1}, \dots, K^{(\text{spec})}_{k}\}$ ranked by cosine similarity between the query embedding and the stored cases. Then we get the \emph{specifications} and \emph{detailed vulnerability analysis} in \emph{top-$k$ cases}.

\textbf{Retrieval Query Construction for domain-specific specifications.} We design a retrieval process that emphasizes identifiers and domain concepts, which capture the most relevant features of a target function likely to correlate with similar exploitation cases. We first use \textbf{Identifier Keywords and Domain Keywords.} For each target function, we instruct the LLM to generate two complementary sets of retrieval keywords. 
The first are \emph{identifier keywords}, which anchor the function to its concrete software context. 
Following a priority hierarchy, LLM generates CPE vendor or product names (e.g., ``ImageMagick''), repository or organization names, and component or module names. 
These identifiers ensure that vulnerabilities from the same project or closely related codebases are prioritized.  
The second are \emph{domain keywords}, which characterize the broader functionality exposed to attackers. 
Here, LLM generates descriptors such as protocol names (e.g., TLS, HTTP, SSH), file formats (e.g., PNG, XML, PDF), or system components (e.g., parser, allocator, validator). 
Such domain keywords capture recurring attack surfaces even across different projects, enabling the discovery of similar exploitation strategies. Then we \textbf{filter and rank the retrieved cases in our comprehensive CVE dataset.} To simulate how historical vulnerabilities would realistically assist detection, we adopt a two-stage filtering and ranking process:
(i) We apply \emph{temporal constraints}. For a target function associated with CVE $x$, we retain only cases whose publication date precedes that of $x$, and whose CVE identifiers were assigned earlier in the official numbering sequence.   
(ii) We perform \emph{keyword-based filtering} using the identifier and domain keywords generated in the previous step. Candidate CVEs are retained only if their CPE fields or CVE descriptions contain at least one of the generated keywords. To avoid overwhelming matches, keywords that return more than 500 CVEs are discarded as overly broad.  
(iii) We apply a \emph{naive ranking scheme} to prioritize the most informative candidates. Each case is scored based on three factors: (a) presence of attacker-related terms, like "attack", (b) description length as a proxy for detail richness, and (c) recency, where more recent vulnerabilities are favored. We compute an averaged weight across these dimensions to produce the final ranking. Using these strategies, we retrieve the most relevant top-$k$ CVE cases in the same domain, 
$\{K^{(\text{CVE})}_{1}, \dots, K^{(\text{CVE})}_{k}\}$, 
which constitute the candidate set after filtering and ranking.

\subsection{Specification-guided Detection}
\label{sec:skb}

In the final phase, VulInstruct integrates retrieved knowledge into the detection pipeline. \begin{change}
The process consists of three sequential stages: (1) knowledge scoring, which filters retrieved candidate by relevance; (2) domain-specific specification generation, which dynamically abstracts rules from filtered CVE cases; and (3) structured reasoning, which integrates all surviving knowledge to produce a vulnerability judgment with explanation.
\end{change}

\textbf{Knowledge Scoring.}
For each target function, we prompt the LLM to evaluate the retrieved \emph{top-$k$ specifications}, their corresponding \emph{detailed vulnerability cases}, and candidate \emph{top-$k$ CVE cases}. LLM scores each specification, detailed case, and CVE case on a 1-10 scale.
We establish a simple 1-10 scoring rubric, where 10 points indicate highly relevant cases with strong alignment in vulnerability type and code features, and lower scores indicate weaker or only partial similarity. This scoring step ensures that subsequent reasoning is guided primarily by the most relevant knowledge rather than by noisy or tangential cases. We then apply a unified \emph{threshold} to filter out low-scoring items. The surviving cases form the \emph{high-relevance pool} of specifications and CVEs that guide reasoning.

\textbf{Domain-specific specifications Generation.} For the filtered set of relevant domain CVEs in \emph{high-relevance pool}. VulInstruct induces reusable specifications by jointly analyzing multiple related cases. Unlike general specifications that are pre-extracted at the patch level, domain-specific specifications are \emph{dynamically abstracted} during detection. 
We leverage information from CVE descriptions, CWE types, and CPE identifiers, together with detailed exploitation notes, to identify recurring exploitation mechanisms. 
Formally, each domain-specific specification is expressed as:
\begin{equation}
\text{AS-\texttt{[DOMAIN]}-\texttt{[ID]}}:\text{[Expected Safe Behavior]}
\end{equation}
Here, \texttt{ID} correspond to the same concepts in Equation~\ref{eq:domain}, \begin{change}\texttt{DOMAIN} is open-ended and determined by the LLM based on the recurring exploitation pattern observed across retrieved CVE cases, and \texttt{AS} stands for Attack-derived Specification.\end{change}
In addition to the expected safe behavior, each entry also records its supporting \emph{recurring exploitation mechanism}, derived by jointly analyzing multiple CVEs (e.g., heap-based buffer overflows from unchecked decoding in CVE-2014-0011 and CVE-2014-9629).  
This ensures that the specification is not only a defensive rule but also explicitly grounded in concrete exploitation patterns repeatedly observed in practice.

\textbf{Structured Reasoning.}
\begin{change}
After scoring and filtering, the surviving knowledge is assembled into the detection prompt. Specifically, we concatenate three types of filtered knowledge under the ``LLM-filtered Security Knowledge'' field. Using the filtered knowledge as context, we instruct the LLM to follow
a structured reasoning process. Here, ``structured'' refers to a
constrained, layered reasoning strategy enforced through prompt design,
rather than a structured intermediate output format. Specifically, our
prompt decomposes the analysis into three progressive layers:
(1)~surface symptom identification, (2)~root cause investigation via
data/control flow and attacker exploitation path tracing, and
(3)~architectural and contextual analysis. The model must produce
findings from all three layers before deriving its final vulnerability
judgment. Throughout this process, the model aligns the target function
against all scored specifications, enabling it to explain vulnerabilities
in terms of established security knowledge rather than relying on shallow
pattern matching.

\end{change}

\section{Experiment Design}\label{sec:experiment}
\subsection{Datasets}
\label{subsec:datasets}

\begin{change}
We clarify the roles of datasets with respect to the two-layer Specification Knowledge Base (Section~\ref{sec:skb}). CORRECT is used to construct the reusable general specification knowledge base 
(Section~\ref{subsec:general_specification_knowledge_base}), while the domain evidence base 
is derived from the NVD/CVE corpus, whose construction has been described in 
Section~\ref{subsec:domain_evidence_base}.
PrimeVul is used solely as the evaluation benchmark.

\textbf{Evaluation Benchmark: PrimeVul}~\cite{ding2024vulnerability} is a widely adopted 
vulnerability detection benchmark. The key feature of PrimeVul is its \textit{temporal data 
splitting}---training data comes from vulnerabilities before a cutoff date, while test data comes 
from after, ensuring a realistic evaluation where detection systems must generalize from historical 
patterns to identify new vulnerabilities.
\begin{change}
Additionally, PrimeVul pairs each vulnerable function with its patched version, requiring models to 
distinguish semantically near-identical code---making the benchmark particularly challenging even for 
state-of-the-art LLMs.
We use the official PrimeVul pair-wise test split (838 samples, 50\% vulnerable / 50\% patched), consistent 
with prior work~\cite{wen-etal-2025-boosting, widyasari2025let}.
\end{change} \textbf{Knowledge Source: CORRECT}~\cite{li2025everything} dataset contains 2000 vulnerable-patched 
function pairs and provides comprehensive contextual information, including (i) complete callee functions 
at multiple depths, (ii) type declarations and data structures, (iii) global variables and 
constants, and (iv) import statements and module dependencies. We leverage this high-quality dataset 
to build our general specification knowledge base.
Specifically, we use 1{,}338 vulnerable--patched function pairs, all disclosed strictly before the 
PrimeVul test set cutoff date, ensuring no temporal overlap with evaluation data. 

\end{change}

\subsection{Baselines}
\label{subsec:baselines}
We compare VulInstruct against state-of-the-art LLM-based vulnerability detection approaches across different paradigms:
\textit{Prompting Methods.}
\textbf{Chain-of-Thought} \cite{ding2024vulnerability}: We follow Ding et al.'s approach, prompting the LLM with step-by-step reasoning for vulnerability analysis. This serves as the baseline of direct prompting without enhancements.
\textit{Fine-tuning Methods.}
\textbf{ReVD}~\cite{wen-etal-2025-boosting} is the current state-of-the-art fine-tuned approach for vulnerability detection. It uses Qwen2.5-Coder as the base model and synthesizes 28,000 vulnerability analysis reasoning data for training, achieving state-of-the-art performance on PrimeVul. 
\textbf{VulTrial}~\cite{widyasari2025let}: An agent-based vulnerability analysis framework that employs four role-specific agents (researcher, author, moderator, and review board). It fine-tunes GPT-4o with role-specific instructions and achieves state-of-the-art performance among agent-based approaches on PrimeVul.
\textit{Agent-based Methods.}
\textbf{GPTLens}~\cite{hu2023large}: multi-agent framework that automates vulnerability analysis workflows and iterative reasoning to enhance LLMs reasoning ability.
\textit{Retrieval-Augmented Methods.}
\textbf{Vul-RAG}~\cite{du2024vul} retrieves relevant vulnerability knowledge based on functional semantics and uses it to augment the detection process. The method extracts multi-dimensional knowledge, including vulnerability causes and fixing solutions.

\subsection{Evaluation Metrics}
\label{subsec:metrics}


As discussed in \ref{relate_work_eval}, using binary classification alone as the evaluation criterion for LLMs has been widely questioned. A model may predict the correct label while relying on spurious code patterns, which do not reflect the actual vulnerability scenario. This raises concerns that accuracy alone cannot demonstrate true understanding of the root cause of vulnerabilities. 

Therefore, we adopt the CORRECT evaluation metrics~\cite{li2025everything}, proposed by Li et al., which assess both the vulnerability label and the underlying reasoning process. LLMs must both predict correctly and provide reasoning that is consistent with ground-truth evidence, including CVE descriptions, patches, and commit messages, as verified by an LLM-as-a-Judge model (i.e., another LLM acting as an evaluator). With the correctness of predictions and reasoning process defined under CORRECT, we compute standard metrics for vulnerability detection, including accuracy $\frac{TP+TN}{TP+TN+FP+FN}$, precision $\frac{TP}{TP+FP}$, recall $\frac{TP}{TP+FN}$, and F1-score $\frac{2 \cdot \text{precision} \cdot \text{recall}}{\text{precision} + \text{recall}}$. In addition, we employ \emph{pair-wise prediction metrics}~\cite{ding2024vulnerability, wen-etal-2025-boosting} to assess whether models can distinguish vulnerable functions from their patched functions. Formally, let $(p_v, p_f) \in \{0,1\}^2$ denote prediction outcomes, where $p_v$ indicates whether the model correctly identifies the \emph{ground-truth vulnerability} in the original function (based on both label and reasoning process), and $p_f$ indicates whether the model correctly recognizes the fixing in the function. In our evaluation, we adopt the commonly used \textbf{P-C} metric, where $(1,0)$ denotes ideal detection: the model correctly identifies the vulnerability in the original code but not in its patched version. In addition, we employ the \textbf{VP-S} score~\cite{wen-etal-2025-boosting}, defined as $\text{VP-S}=\text{P-C}-\text{P-R}$, where P-R $(0,1)$ represents reversed prediction. VP-S thus provides a stricter measure of pair-wise discrimination by rewarding correct detections while penalizing such reversed errors.
\begin{change}
We further introduce \textbf{Unique}, which measures the proportion of vulnerabilities exclusively detected by a given method among all successfully detected CVEs. Let $\mathcal{C}$ denote the set of CVEs detected by at least one method and $\mathcal{C}_m$ the subset detected by method $m$. The Unique score is defined as $\text{Unique}(m)=\frac{|{c \in \mathcal{C}m \mid c \notin \mathcal{C}{m'},\ \forall m' \neq m}|}{|\mathcal{C}|}$. This metric reflects a method’s complementary detection capability by quantifying vulnerabilities missed by all other systems.
\end{change}
For clarity, we provide a section in the supplementary material (Section A in the Appendix file) that details the formulation in CORRECT evaluation framework.

\subsection{Implementation Details}
We provide key implementation details for reproducibility: (i) We use Qwen3-Embedding-0.6B to generate embeddings for retrieval queries and stored system identification and functional semantics. (ii) For both types of security specifications, we retrieve top-$k=10$ candidates initially. In knowledge scoring, we apply a unified threshold of $\geq 6$ points (6, 6, 6) to filter three types of knowledge. (iii) For knowledge extraction in two type specification base, keyword generation, knowledge scoring, we use DeepSeek-V3 model. (iv)
For CORRECT evaluation metric, we use GPT-5 as LLM-as-a-Judge model, which represents the current state-of-the-art in evaluation capabilities, replacing the GPT-4o used in the original CORRECT implementation. (v) Details of our prompt formats in extracting specifications, knowledge scoring, and vulnerability detection can be found in the supplementary material (Section B in the Appendix file). We also provide the technical details about our automatic tool in code context extraction (Section F in the Appendix file).


\section{Results and Analysis}\label{sec:results}

\subsection{RQ1: How effective is VulInstruct in vulnerability detection compared to state-of-the-art approaches?}
\label{sec:rq1}

\begin{table*}[h]
\vspace{-0.4cm}
\caption{Vulnerability detection performance on PrimeVul. Best in bold; second-best underlined.}
\setlength{\tabcolsep}{2pt}
\vspace{-0.3cm}
\label{tab:correct_results}
\centering
\resizebox{0.95\linewidth}{!}{
\begin{tabular}{@{}llcccccc|cc@{}}
\toprule
& & & \multicolumn{5}{c|}{\textbf{Standard (\%)}} & \multicolumn{2}{c}{\textbf{Pairwise (\%)}} \\
\textbf{Method} & \textbf{Model} & \textbf{Trained} & Acc.$\uparrow$ & Prec.$\uparrow$ & Rec.$\uparrow$ & Unique$\uparrow$ & F1-Score$\uparrow$ &  P-C$\uparrow$ & VP-S$\uparrow$ \\
\midrule
\multicolumn{10}{l}{\textit{Prompting-based Methods}} \\
Ding et al.(CoT) & Deepseek-V3 & \xmark & 49.9 & 49.5 & 12.2 & 1.8 & 19.5 & 3.9 & 1.4 \\
\midrule
\multicolumn{10}{l}{\textit{Fine-tuning Methods}} \\
ReVD & Qwen2.5-Coder & \checkmark & 49.5 & 48.0 & 11.5 & 5.8 & 18.6 & 7.4 & $-$0.9 \\
VulTrial & GPT-4o & \checkmark & \underline{53.4} & \underline{59.9} & 20.8 & \underline{10.2} & 30.9 & 12.3 & \underline{6.9} \\
\midrule
\multicolumn{10}{l}{\textit{Agent-based Methods}} \\
VulTrial & DeepSeek-V3 & \xmark & 51.9 & 57.4 & 10.0 & 10.2 & 17.0 & 6.6 & 2.7 \\
GPTLens & DeepSeek-V3 & \xmark & 51.3 & 52.8 & \underline{25.0} & 4.4 & \underline{33.9} & \underline{13.0} & 2.7 \\
\midrule
\multicolumn{10}{l}{\textit{Retrieval-Augmented Methods}} \\
Vul-RAG & DeepSeek-V3 & \xmark & 50.5 & 58.3 & 3.4 & 0.4 & 6.5 & 2.5 & 1.0 \\
\textbf{VulInstruct} & DeepSeek-V3 & \xmark & \textbf{53.9} & 55.8 & \textbf{37.7} & \textbf{24.3} & \textbf{45.0} & \textbf{17.2} & \textbf{7.4} \\
\rowcolor{gray!10}
\multicolumn{3}{l}{\textit{Relative Improvement over best baseline}} & +0.9\% & -6.8\% & +50.8\% & +138.2\% & +32.7\% & +32.3\% & +7.2\% \\
\bottomrule
\end{tabular}}
\end{table*}

Table~\ref{tab:correct_results} shows the results of vulnerability detection on the PrimeVul dataset. \textbf{VulInstruct achieves state-of-the-art performance with 45.0\% F1 score, surpassing GPTLens by 32.7\%.} More critically, VulInstruct achieves 37.7\% recall, an improvement of 50.8\% over GPTLens, demonstrating the superior ability to discover vulnerabilities. While fine-tuned VulTrial with GPT-4o achieves higher precision (59.9\%), its recall remains limited at 20.8\%, reflecting a high-confidence but low-coverage detection strategy. In contrast, VulInstruct maintains balanced performance with 55.8\% precision while achieving 37.7\% recall, demonstrating that VulInstruct can identify more vulnerabilities without sacrificing accuracy. Furthermore, we introduce the Unique metric which represents the percentage of vulnerabilities exclusively detected by that method among all 226 successfully detected CVEs across all systems. VulInstruct achieves 24.3\% uniqueness, significantly outperforming all baselines. Notably, we analyze Vul-RAG and ReVD failures in Appendix C, revealing why retrieval-based and reasoning-enhanced approaches underperform.

\textbf{VulInstruct also demonstrates superior capability in distinguishing vulnerable from patched code.} In pairwise evaluation, VulInstruct achieves 17.2\% P-C accuracy and 7.4\% VP-S score, improving by 32.3\% over the best baseline. This indicates our approach not only identifies vulnerabilities but understands how patches address underlying security issues.

\subsection{RQ2: What is the contribution of each module in VulInstruct?}
\label{sec:rq2}

\begin{table}[h]
\caption{Ablation study and knowledge utilization in VulInstruct. The top part shows performance by subsets of test samples, grouped by retrieved knowledge sources (pair-wise metrics are omitted since subsets are not directly comparable). 
The bottom part reports ablation results where one source is entirely removed. }

\label{tab:ablation_knowledge}
\centering
\resizebox{0.85\linewidth}{!}{
\begin{tabular}{@{}lccccccc@{}}
\toprule
\textbf{Variant / Subset} & \textbf{Ratio (\%)} & \textbf{Acc.} & \textbf{Prec.} & \textbf{Rec.} & \textbf{F1} & \textbf{P-C} & \textbf{VP-S} \\
\midrule
\textbf{Full VulInstruct (Deepseek-V3)} & 100.0 & 53.9 & 55.8 & 37.7 & 45.0 & 17.2 & 7.4 \\
\midrule
\quad General + Domain-specific        & 73.5 & 55.5 & 58.4 & 41.2 & 48.3 & -- & -- \\
\quad General only   & 25.7 & 52.1 & 51.4 & 36.2 & 42.5 & -- & -- \\
\quad No knowledge matched & 0.7  & 50.0 & 0.0  & 0.0  & 0.0  & -- & -- \\
\quad Domain-specific only             & 0.1  & 100.0 & 0.0  & 0.0  & 0.0  & -- & -- \\
\midrule
w/o Domain-specific specifications          & --   & 50.6 & 50.9 & 33.4 & 40.4 & 14.2 & 1.0 \\
w/o General specifications  & --   & 51.7 & 52.6 & 33.7 & 41.0 & 16.2 & 3.7 \\
\changetable{w/o Dynamic specification generation} & \changetable{--} & \changetable{48.9} & \changetable{48.2} & \changetable{29.4} & \changetable{36.5} & \changetable{11.9} & \changetable{-0.9} \\
\bottomrule
\end{tabular}}
\vspace{-0.4cm}
\end{table}

To understand how different knowledge sources contribute to detection, we analyze VulInstruct under two complementary perspectives: knowledge utilization patterns in the full model and ablation studies where one source is entirely removed.

\textbf{Knowledge utilization.} For each input, 
VulInstruct retrieves 10 relevant general knowledge (general specifications and detailed vulnerability cases) from General Specification Knowledge Base and 10 relevant CVE cases from the Domain Evidence Base. 
After Knowledge scoring, each test sample in vulnerability detection falls into one of four mutually exclusive categories: (i) both general knowledge and Domain-specific specifications left, (ii) only general knowledge left, (iii) only domain-specific specifications left, or (iv) no knowledge left. As shown in Table~\ref{tab:ablation_knowledge}, most samples (73.5\%) belong to case (i), achieving the highest F1-score of 48.3\%. Even when only general specifications or vulnerability detailed cases are left (25.7\%), performance remains strong at 42.5\% F1. In contrast, cases with only domain-specific specifications (0.1\%) or no matched knowledge (0.7\%) yield near-zero performance. These results indicate that two types of specifications are complementary: general specifications provide the essential detection capability, while domain-specific specifications enhance it when combined. For most samples, VulInstruct can retrieve at least one type of relevant knowledge, enabling more effective and reliable vulnerability detection.

\textbf{Ablation.} To further isolate contributions, we conduct ablation studies with entire knowledge sources removed. Removing domain-specific specifications causes F1-score to drop from 45.0\% to 40.4\% ($-$4.6\%), while removing general specifications and detailed vulnerability cases reduces it to 41.0\% ($-$4.0\%). Interestingly, while domain-specific specifications appear in fewer samples (73.5\% vs. 99.2\% for general specifications and detailed vulnerability cases), their contribution to performance is slightly larger. This suggests that domain-specific specifications, when applicable, provide highly valuable domain-specific exploitation knowledge that complements the broader security knowledge captured in general knowledge. The VP-S metric shows a more dramatic difference: dropping from 7.4\% to 1.0\% without domain-specific specifications versus 3.7\% without general knowledge, indicating its importance for distinguishing vulnerable from patched code.
\begin{change}
We further evaluate the contribution of dynamic specification generation.
Instead of generating domain-specific specifications, we directly provide the retrieved CVE case descriptions to the model. This change reduces F1-score from 45.0\% to 36.5\%, showing that dynamic abstraction is essential. This result suggests that raw CVE descriptions may contain overly detailed or highly specific 
contextual information, whereas dynamic abstraction helps distill reusable security constraints, 
enabling better generalization across different vulnerability instances.

\begin{figure}[t]
    \centering
    \includegraphics[width=0.95\textwidth]{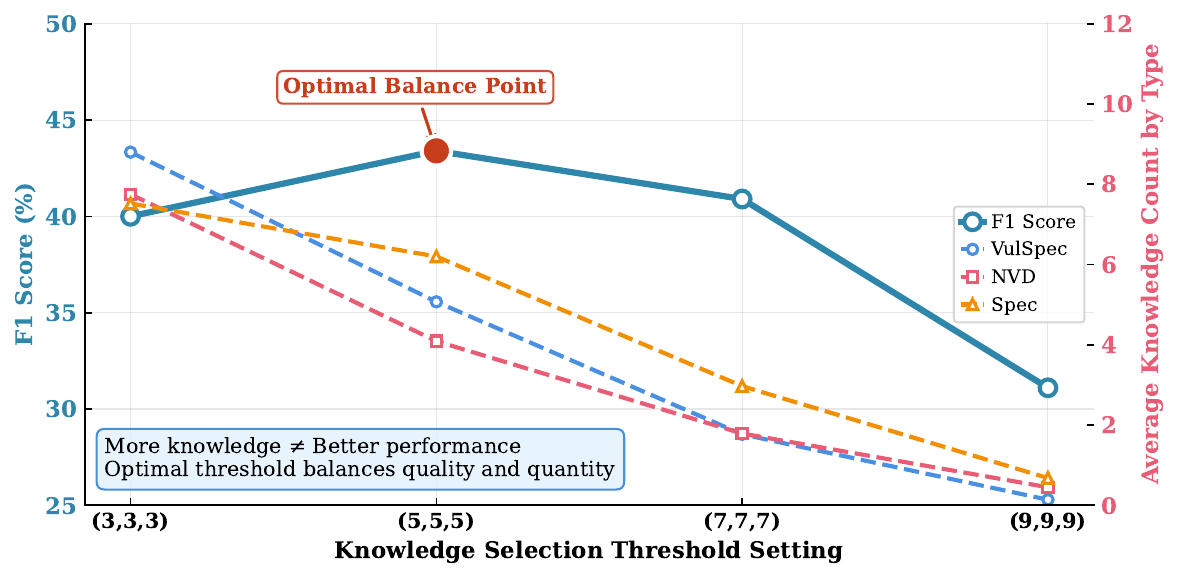}
    \vspace{-0.3cm}
    \caption{Knowledge Selection Threshold Ablation: The inverted-U relationship in RAG-based vulnerability detection. \textit{Spec} represents general specifications, \textit{VulSpec} denotes corresponding detailed vulnerability cases, and \textit{NVD} indicates CVE cases (which are subsequently transformed into domain-specific specifications). }
    \label{fig:knowledge-threshold}
    \vspace{-0.5cm}
\end{figure}

\textbf{Threshold and Sensitivity Analysis.}
We study how the knowledge selection threshold and the top-$K$ retrieval
setting affect detection performance.  To control experimental cost, we
randomly sample 100 out of 435 vulnerability–patch pairs and vary one
parameter at a time.

\begin{table}[h]
\changetable{
\caption{
Sensitivity analysis on knowledge selection threshold and
top-$K$ retrieval setting. All experiments use 100 randomly
sampled vulnerability--patch pairs.
}
\vspace{-0.2cm}
\label{tab:sensitivity_compact}
\centering
\small
\setlength{\tabcolsep}{4.5pt}
\begin{tabular}{l ccccccc | cccc}
\toprule
& \multicolumn{7}{c|}{\textbf{Threshold (Spec, VulSpec, NVD)}} & \multicolumn{4}{c}{\textbf{Top-$K$}} \\
& (6,6,6)$^{\dagger}$ & (3,6,6) & (9,6,6) & (6,3,6) & (6,9,6) & (6,6,3) & (6,6,9) & 3 & 5 & 7 & 10$^{\dagger}$ \\
\midrule
F1 (\%) & 37.5 & 38.5 & 37.2 & 37.0 & 27.5 & 38.1 & 34.2 & 34.2 & 35.9 & 33.3 & 39.5 \\
\bottomrule
\multicolumn{12}{l}{\footnotesize $^{\dagger}$ default setting.}
\end{tabular}
}
\vspace{-0.2cm}
\end{table}

\end{change}

\textbf{Unified threshold.}
Figure~\ref{fig:knowledge-threshold} varies the threshold uniformly
across all three knowledge types.  F1-score follows an inverted-U curve:
a lenient threshold (3,3,3) preserves more knowledge but introduces
noise (40.2\% F1); an aggressive threshold (9,9,9) retains on average
fewer than two items per type and performance drops sharply (31.1\% F1).
The peak occurs at (5,5,5) with 43.4\% F1, confirming that an
intermediate threshold best balances knowledge quality and coverage.
\begin{change}
\textbf{Per-type threshold sensitivity.}
Table~\ref{tab:sensitivity_compact} further isolates the effect
of each knowledge type by varying one threshold while fixing the others
at~6.  The default setting (6,6,6) achieves 37.5\% F1.  Loosening a
single threshold to~3 has limited impact (at most +1.0\% F1), whereas
tightening to~9 causes notable drops—particularly for VulSpec
($-$10.0\%) and NVD ($-$3.3\%)—indicating that domain-specific evidence
is especially sensitive to over-filtering.
\textbf{Top-$K$ retrieval.}
Table~\ref{tab:sensitivity_compact} examines the number of
retrieved candidates per knowledge type.  Performance generally improves
from top-3 (34.2\%) to top-10 (39.5\%), as a larger candidate pool
increases the chance of surfacing highly relevant specifications.  The
minor fluctuation at top-7 (33.3\%) is attributable to sampling variance
on the 100-pair subset and does not alter the overall upward trend. Overall, VulInstruct is reasonably stable across a range of settings.
Our default configuration strikes a practical
balance between knowledge quality and coverage, achieving competitive
performance.

\end{change}

\subsection{RQ3: How well does VulInstruct generalize across different vulnerability types and models?}
\label{sec:rq3}

Beyond overall performance, we examine VulInstruct's generalization across diverse vulnerability types and LLMs. 
Following the strict CORRECT evaluation, we define \textbf{MATCH rate} as the percentage of vulnerable samples where the model correctly identifies both the label and the root cause: 
MATCH rate = (correctly identified root causes) / (total vulnerable samples). This metric is important because simply predicting a ``vulnerable'' label does not ensure practical usefulness—security analysts need to know \textit{why} the code is vulnerable. 
MATCH rate thus directly reflects the model’s ability to discover and explain vulnerabilities rather than relying on superficial correlations.

\begin{figure}[htbp]
    \centering
    \begin{subfigure}[t]{0.44\linewidth}
        \centering
        \includegraphics[width=\linewidth]{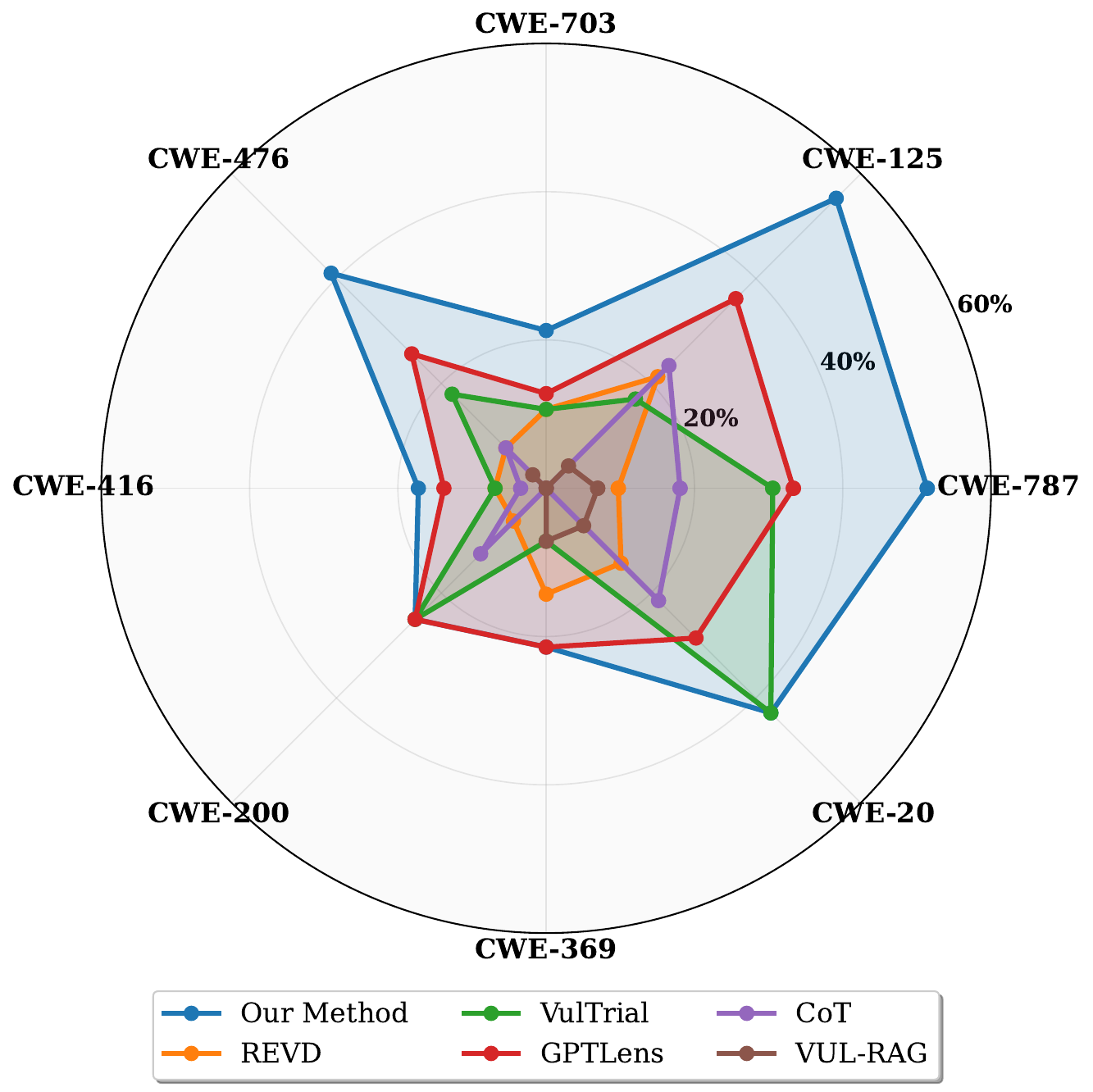}
        \caption{CWE Radar Chart}
        \label{fig:cwe_radar}
    \end{subfigure}
    \hfill
    \begin{subfigure}[t]{0.54\linewidth}
        \centering
        \includegraphics[width=\linewidth]{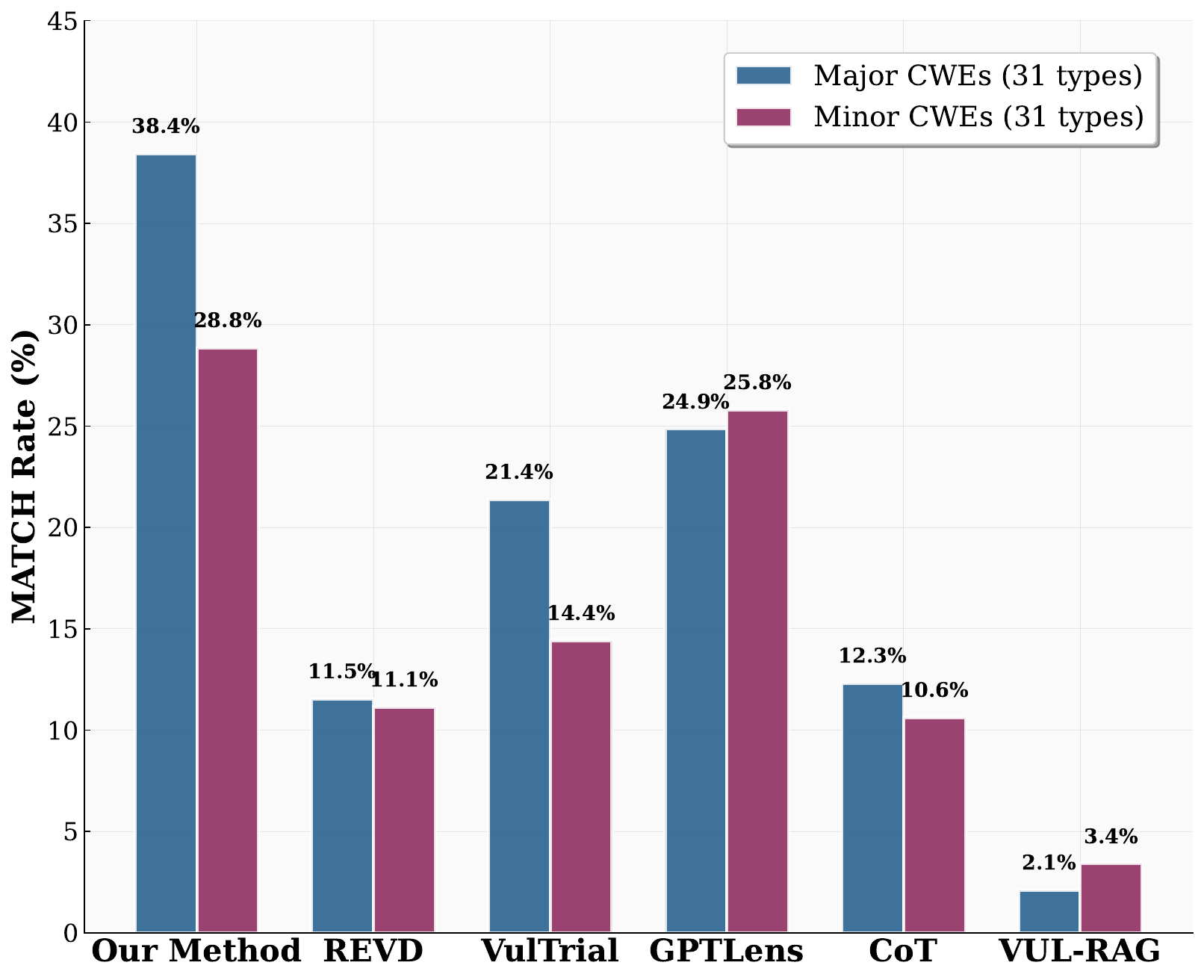}
        \caption{Head/Tail Performance}
        \label{fig:head_tail}
    \end{subfigure}
    \vspace{-0.3cm}
    \caption{Comparison of two experimental results.}
    \label{fig:two_results}
    \vspace{-0.3cm}
\end{figure}

\textbf{Across CWE types.} Figure~\ref{fig:cwe_radar} presents a radar chart comparing MATCH rates over the eight most frequent CWE categories in PrimeVul. VulInstruct consistently outperforms all baselines across diverse vulnerability types, including memory safety issues (e.g., CWE-125, CWE-787) and resource management flaws (e.g., CWE-416, CWE-476). While competing methods such as VulTrial and GPTLens exhibit strong performance only in certain categories, our approach achieves more balanced coverage. \textbf{Head vs.\ tail performance.} 
We evaluate performance on the long-tail distribution of vulnerability types, by ranking all 62 CWE types by their sample count in descending order. We then partition them into two equal groups: the top 31 CWE types (major/head group which contains 91.7\% of all vulnerable samples) and the bottom 31 CWE types (minor/tail group with the left 8.3\%). Figure~\ref{fig:head_tail} compares detection on major (head) CWEs and rare (tail) CWEs. VulInstruct attains the highest MATCH rate on both groups (38.4\% on major CWEs and 28.8\% on minor CWEs), outperforming GPTLens (25.8\% and 24.9\%) and VulTrial (21.4\% and 14.4\%). These results indicate that VulInstruct not only improves detection of common vulnerabilities but also alleviates the long-tail challenge by transferring knowledge to rare vulnerability types.

\begin{change}

\textbf{Across benchmarks.} Table~\ref{tab:generalize_sven} shows that VulInstruct generalizes effectively across independent vulnerability benchmarks, consistently outperforming representative baselines on the SVEN dataset \cite{he2023large}. 
SVEN contains diverse real-world vulnerabilities constructed independently from PrimeVul, allowing us to assess cross-benchmark transferability. 
We evaluate all C/C++ cases in SVEN, comprising 417 vulnerable functions and their corresponding vulnerable–patch pairs, while reusing the same experimental setup, including CORRECT as the knowledge source and GPT-5 as the LLM-as-a-Judge evaluator.

\begin{table*}[t]
\changetable{
\vspace{-0.35cm}
\caption{Generalization performance on SVEN.}
\vspace{-0.25cm}
\label{tab:generalize_sven}
\centering
\setlength{\tabcolsep}{3pt}
\renewcommand{\arraystretch}{1.05}
\resizebox{0.84\linewidth}{!}{
\begin{tabular}{llcccc|ccc}
\toprule
\textbf{Type} & \textbf{Method} 
& \multicolumn{4}{c|}{\textbf{Standard (\%)}} 
& \multicolumn{3}{c}{\textbf{Pairwise (\%)}} \\
\cmidrule(lr){3-6} \cmidrule(lr){7-9}
& 
& Acc.$\uparrow$ & Prec.$\uparrow$ & Rec.$\uparrow$ & F1$\uparrow$
& P-C$\uparrow$ & P-R$\downarrow$ & VP-S$\uparrow$ \\
\midrule

Fine-tuning-based & VulTrial
& 51.2 & 60.0 & 7.3 & 13.1 
& 6.8 & 4.3 & 2.4 \\

Agent-based & GPTLens
& 55.3 & 64.9 & 23.1 & 34.1 
& 18.8 & 8.2 & 10.6 \\

Retrieval-based & Vul-RAG
& 49.2 & 47.9 & 18.5 & 26.7
& 15.5 & 17.1 & $-$1.6 \\

Retrieval-based & VulInstruct
& \textbf{57.1} & 57.0 & \textbf{57.6} & \textbf{57.3} 
& \textbf{24.7} & 10.6 & \textbf{14.1} \\

\bottomrule
\end{tabular}}
}
\vspace{-0.25cm}
\end{table*}

VulInstruct achieves the best overall performance with an F1-score of 57.3\% and a recall of 57.6\%, substantially surpassing GPTLens (34.1\% F1) and Vul-RAG (26.7\% F1). 
The large recall improvement indicates stronger capability in identifying vulnerable instances rather than relying on conservative predictions. 
This advantage also transfers to pairwise evaluation, where VulInstruct attains the highest Patch-Correctness (P-C) and Vulnerability-Preserving Score (VP-S), suggesting more reliable reasoning about vulnerability causes and patch effectiveness across benchmark distributions.
\end{change}

\textbf{Across models.} Table~\ref{tab:model_comparison} shows that \textbf{VulInstruct consistently improves different families of LLMs compared with their CoT approach}
On GPT-OSS-120B, VulInstruct boosts recall by +72.9\% and F1 by +45.4\%, complementing its already strong precision. 
On Claude-Sonnet-4, the relative gains are even more striking: VP-S more than triples (from 2.2\% to 7.3\%) and P-C more than doubles, 
demonstrating that VulInstruct effectively enhances models with weaker intrinsic security awareness. 
Finally, DeepSeek-R1 with VulInstruct shows the strongest overall results, achieving the best P-C (22.8\%), accuracy (56.6\%), 
and precision (62.8\%), establishing a new state of the art.

\begin{table*}[htbp]
\vspace{-0.3cm}
\caption{Performance of VulInstruct across different language models on PrimeVul dataset. Improvement row represents relative improvement (\%)}
\vspace{-0.1cm}
\setlength{\tabcolsep}{2pt}
\label{tab:model_comparison}
\centering
\resizebox{0.8\linewidth}{!}{
\begin{tabular}{@{}llcccccc@{}}
\toprule
\textbf{Model} & \textbf{Method} & \textbf{Acc. (\%)} & \textbf{Prec. (\%)} & \textbf{Rec. (\%)} & \textbf{F1 (\%)} & \textbf{P-C (\%)}  & \textbf{VP-S (\%)}\\
\midrule
\multirow{3}{*}{GPT-OSS-120B} 
 & Baseline & 55.1 & \textbf{70.5} & 17.7 & 28.2 & 15.9 & 10.1\\
 & + VulInstruct & \textbf{56.1} & 62.4 & \textbf{30.6} & \textbf{41.0} & \textbf{17.6} & \textbf{11.8} \\
 & \textit{Improvement} 
   & \textit{\textcolor{red}{+1.8\% ↑}} 
   & \textit{\textcolor{green}{-11.5\% ↓}} 
   & \textit{\textcolor{red}{+72.9\% ↑}} 
   & \textit{\textcolor{red}{+45.4\% ↑}} 
   & \textit{\textcolor{red}{+10.7\% ↑}} 
   & \textit{\textcolor{red}{+16.8\% ↑}} \\
\midrule
\multirow{3}{*}{Claude-Sonnet-4} 
 & Baseline & 51.3 & 52.1 & 32.0 & 39.6 & 3.2 & 2.2\\
 & + VulInstruct & \textbf{53.6} & \textbf{55.6} & \textbf{35.8} & \textbf{43.5} & \textbf{6.9} & \textbf{7.3}\\
 & \textit{Improvement} 
   & \textit{\textcolor{red}{+4.5\% ↑}} 
   & \textit{\textcolor{red}{+6.7\% ↑}} 
   & \textit{\textcolor{red}{+11.9\% ↑}} 
   & \textit{\textcolor{red}{+9.8\% ↑}} 
   & \textit{\textcolor{red}{+115.6\% ↑}} 
   & \textit{\textcolor{red}{+231.8\% ↑}} \\
\midrule
\multirow{3}{*}{DeepSeek-R1} 
 & Baseline & 53.7 & 59.2 & 23.9 & 34.0 & 18.9 & 4.4 \\
 & + VulInstruct & \textbf{56.6} & \textbf{62.8} & \textbf{32.2} & \textbf{42.6} & \textbf{22.8} & \textbf{13.7} \\
 & \textit{Improvement} 
   & \textit{\textcolor{red}{+5.4\% ↑}} 
   & \textit{\textcolor{red}{+6.1\% ↑}} 
   & \textit{\textcolor{red}{+34.7\% ↑}} 
   & \textit{\textcolor{red}{+25.3\% ↑}} 
   & \textit{\textcolor{red}{+20.6\% ↑}} 
   & \textit{\textcolor{red}{+211.4\% ↑}} \\
\bottomrule
\end{tabular}}
\vspace{-0.3cm}
\end{table*}

\subsection{RQ4: What makes VulInstruct more effective than existing approaches?}
\label{sec:rq4}

\begin{figure}[htbp]
    \centering
    \includegraphics[width=0.98\textwidth]{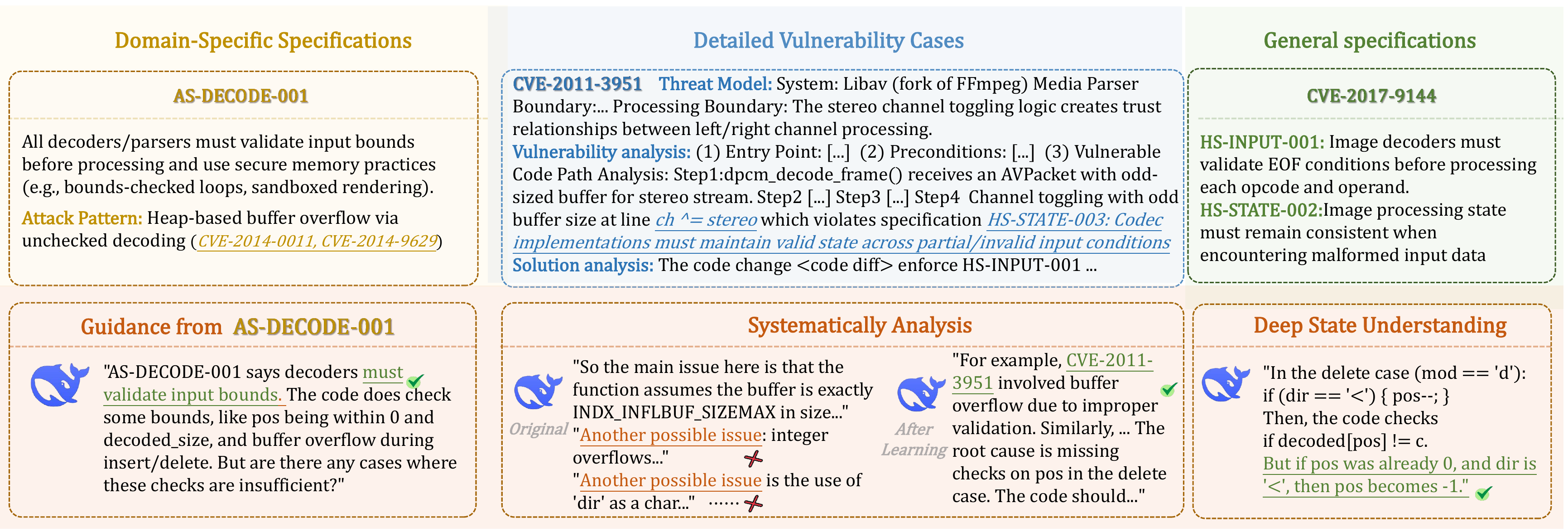}
    \caption{Case Study in detecting CVE-2022-1533}
    \label{fig:case-study}
    \vspace{-0.6cm}
\end{figure}

We analyze VulInstruct’s specification-guided approach using DeepSeek-R1. 
Overall, VulInstruct achieves a 16.7\% improvement in detecting actual vulnerabilities (FN→TP) and an 8.0\% reduction in false positives (FP→TN). 
To better understand how these improvements are achieved, we conducted manual analysis of successful cases, with detailed results provided in Appendix file Section~D of the supplementary material. 
Here, we present one representative
\textbf{case study}: CVE-2022-1533 is a buffer overflow vulnerability in a media decoder, where insufficient input validation on stereo DPCM audio streams leads to memory corruption during decoding. During detection, VulInstruct retained two general specifications and two detailed vulnerability cases from the general specification knowledge base. In addition, six relevant CVE cases were retrieved to induce three domain-specific specifications. Figure~\ref{fig:case-study} highlights three types of knowledge. In particular, the domain-specific specification \texttt{AS-DECODE-001} and the detailed analysis of \texttt{HS-STATE} from CVE-2011-3951 significantly improved DeepSeek-R1’s reasoning. \texttt{AS-DECODE-001} provided the basic guidance that all decoders must validate input bounds, while the ``channel toggling’’ pattern from CVE-2011-3951 (\texttt{ch \^{}= stereo}) inspired the model to further track the interaction of \texttt{dir} and \texttt{pos}, revealing state changes that could trigger the overflow. 
By focusing on \texttt{pos} in depth, the model eventually reached a correct and faithful explanation.

\begin{change}

\subsection{RQ5: What are the computational overhead and scalability characteristics of VulInstruct?}
\label{sec:rq5}

VulInstruct adopts a multi-stage pipeline (context understanding, dual retrieval, knowledge scoring, dynamic abstraction in domain-specific specifications, and final reasoning), which may introduce non-trivial computational overhead. In this RQ, we quantify the end-to-end latency and token cost, and further discuss acceleration strategies to improve practical scalability opportunities.

Table~\ref{tab:cost} reports the per-case cost breakdown of VulInstruct. With sequential execution, VulInstruct takes 150.61\,s per case. The dominant overhead comes from (i) retrieving domain cases (Domain Evidence Retrieval, 45.34\,s) and (ii) LLM-based knowledge scoring (55.27\,s), while other stages (e.g., Context Understanding and Vulnerability Detection) are comparatively smaller.
We also compare against representative baselines. VulInstruct consumes substantially fewer tokens than Vul-RAG (18,432.81 vs. 39,369 input tokens; 3,219.52 vs. 6,593 output tokens), as VulInstruct avoids iterative matching and instead performs a single-round relevance scoring over retrieved candidates. Compared with LangChain-based VulTrial, VulInstruct achieves lower latency (150.61\,s vs. 197.35\,s) with a simpler inference workflow.

\begin{table}[t]
\caption{Computational cost analysis of VulInstruct. The top part compares per-sample cost against baselines. The middle part shows the stage-level breakdown. The bottom part reports practical acceleration results.}
\vspace{-0.2cm}
\label{tab:cost}
\centering
\resizebox{0.85\linewidth}{!}{
\begin{tabular}{@{}lccc@{}}
\toprule
\textbf{Approach / Stage} & \textbf{Input Tokens} & \textbf{Output Tokens} & \textbf{Time (s)} \\
\midrule
Vul-RAG & 39,369 & 6,593 & 115.1 \\
VulTrial & 10,425 & 1,753 & 197.4 \\
\textbf{VulInstruct} & \textbf{18,433} & \textbf{3,220} & \textbf{150.6} \\
\midrule
\quad Context Understanding & 2,444 & 246 & 7.3 \\
\quad Embedding Retrieval & -- & -- & 1.1 \\
\quad Identifier Generation & 2,268 & 94 & 6.7 \\
\quad Domain Evidence Retrieval & -- & -- & 45.3 \\
\quad LLM-based Knowledge Scoring & 9,223 & 1,687 & 55.3 \\
\quad Domain-specific Spec.\ Abstraction & 391 & 490 & 13.6 \\
\quad Vulnerability Detection & 4,107 & 703 & 21.2 \\
\midrule
\changetable{VulInstruct (+ parallel scoring)} & \changetable{--} & \changetable{--} & \changetable{117.3} \\
\changetable{\quad LLM-based Knowledge Scoring (parallel)} & \changetable{--} & \changetable{--} & \changetable{21.9} \\
\changetable{VulInstruct (+ inverted index)} & \changetable{--} & \changetable{--} & \changetable{106.9} \\
\changetable{\quad Domain Evidence Retrieval (indexed)} & \changetable{--} & \changetable{--} & \changetable{1.6} \\
\bottomrule
\end{tabular}}
\vspace{-0.4cm}
\end{table}

\textbf{Scalability opportunities.} Although the sequential pipeline costs 150.61\,s per case, we do not consider this cost fixed. Instead, we view the current implementation as an initial realization, with room for development optimization while preserving the overall architecture.

To examine this potential, we conduct a lightweight parallelization experiment without modifying the pipeline structure. Specifically, we parallelize the LLM-based knowledge scoring stage across retrieved candidates using multi-process execution (5 workers). Under this setting, the per-case latency decreases from 150.6\,s to 117.3\,s. We further investigate optimization opportunities for the domain evidence retrieval stage while preserving the algorithmic workflow. The original system scans temporally filtered CVE descriptions and performs keyword matching at query time, which costs noticeable latency. We instead build a token-level inverted index over CVE descriptions, enabling direct keyword-to-case lookup while keeping the retrieval criteria, filtering rules, ranking strategy described in Section~\ref{sec:skb}.
With indexed retrieval, the domain evidence retrieval time decreases from 45.3\,s to 1.6\,s per case. 
As a result, the overall latency of VulInstruct drops from 150.6\,s to 106.9\,s per case, yielding a 29.0\% end-to-end speedup without affecting detection outcomes.

\end{change}

\section{Discussion}\label{sec:discussion}
\begin{change}
\textbf{General specifications with system context.}
Our observations suggest that the effectiveness of specifications depends on how they are grounded in the target system. In VulInstruct, general specifications are not applied as standalone rules, but are interpreted together with system-level context (e.g., system understanding or threat model). This allows the model to assess both the relevance of a specification and the concrete meaning of its violation in a given setting. For instance, specifications that may appear conflicting in isolation (e.g., emphasizing strict validation versus non-blocking processing) correspond to different system assumptions and attack surfaces, and are naturally disambiguated once contextualized. We therefore view specification-guided detection as most effective when coupled with system understanding, as this combination helps the model move beyond surface patterns toward more consistent identification of vulnerability causes.
\end{change}

\begin{change}

\textbf{Quality of the specification knowledge base.} 
A key concern is whether specifications extracted from different patches for the same CWE may produce contradictory or redundant rules. To investigate this, we manually analyzed 60 specifications extracted under CWE-20 (Improper Input Validation)---one of the most broadly defined CWE categories where contradictions would be most likely. We found zero direct logical contradictions and only one pair with apparent tension, which is naturally resolved by VulInstruct's retrieval and scoring mechanism. We further conducted a diversity analysis showing that these specifications span multiple security domains beyond input validation. A detailed case-by-case analysis is provided in Appendix~H.

\end{change}

\begin{change}
\textbf{Limitation analysis of VulInstruct.} To understand VulInstruct's failure modes, we manually compared DeepSeek-R1 with and without VulInstruct, observing 33 additional true positives but also 11 additional false positives. The majority of these false positives arise when the model correctly identifies the vulnerable version but fails to recognize that the patched version has resolved the issue. This pattern is most pronounced for highly domain-specific patches (e.g., LJPEG decoder logic), suggesting a capability boundary of current LLMs: specifications can successfully guide the model to locate vulnerability patterns, but the model may lack sufficient domain knowledge to verify whether a framework-specific fix truly eliminates the flaw.
\end{change}

\textbf{Potential for Real-World Vulnerability Discovery.}
We further investigate whether our approach can assist in uncovering real-world vulnerabilities. We conducted a preliminary study using our specification-guided vulnerability auditing framework. We simulate a manual vulnerability auditing workflow: Given the specification associated with a known vulnerability and vulnerable function, we construct an agent that automatically generates a small set of git grep commands to retrieve potentially relevant code segments for further inspection. The retrieved contexts are then analyzed by the agent under the guidance of the specification. Through this process, we successfully identified a high-severity logic vulnerability (\begin{change}later assigned CVE-2025-56538\end{change}), which is essentially a variant of the original vulnerability and violates the same specification. We responsibly reported the vulnerability, and the issue was subsequently confirmed by developers, validated via integration testing, and patched. Details are in our supplementary material (Section E in the Appendix file).

Nevertheless, we believe there remain at least two deeper directions to explore based on our work:
\textbf{Richer use of NVD resources.} Current retrieval from NVD is still under-exploited. Each NVD entry typically contains a dynamically updated set of references, including such as patch lists, issue discussions, third-party advisories, developer-provided proof-of-concepts (PoCs), and crash reports. Such information could enrich the model’s reasoning by offering concrete entry points and complete exploit paths, thereby enabling a more comprehensive understanding of vulnerabilities.
\textbf{Improved contextual datasets.} We excluded PrimeVul's training dataset as its early version lacked CVE/CWE mappings and sufficient code context. This risks reducing understanding to single code parts within functions, insufficient for real-world vulnerability reasoning.
Richer contextual datasets would significantly improve the quality of specification knowledge base.

\section{Threats to Validity}\label{sec:threat}

\textbf{Implementation validity.}
We did not identify apparent implementation errors in our study. 
To further validate this, we conducted targeted failure analyses of two representative baselines, Vul-RAG and ReVD. For Vul-RAG, we observed that its retrieval-based decision mechanism tends to prematurely classify samples as secure once a patch match is detected, which leads to a large number of missed vulnerabilities. 
For ReVD, we manually examined 50 vulnerable cases and found that reasoning failures often arise from zero reasoning, incorrect vulnerability categorization, incomplete mechanism understanding, or over-generalization. Detailed analyses are provided in our supplementary material (Section C in the Appendix file).  

\textbf{Experimental fairness.} \textit{The design choice of CWE.} Prior works like CORRECT and Vul-RAG include CWE types as input during detection, which simplifies the task to CWE-specific vulnerability identification. While this is fair within their experimental setup (all compared methods receive CWE information), it does not reflect realistic scenarios where vulnerability types are unknown. In our evaluation, we exclude CWE information from all methods, including VulInstruct and all baselines to ensure both experimental fairness and practical relevance. \textit{The design choice of Dataset.} Furthermore, we did not select the SVEN \cite{he2023large} dataset, which is also a widely adopted high-quality dataset, for evaluation as it lacks the CVE-related metadata required by our evaluation framework.

\textbf{Temporal validity.} To prevent data leakage and ensure realistic evaluation, we enforce strict temporal constraints in our knowledge retrieval. For domain-specific specifications, we only retrieve CVEs whose publication dates and ID assignments precede the target vulnerability in test set. For General specifications, we verify all samples predate PrimeVul's test set cutoff. This temporal ordering ensures we simulate learning from historical vulnerabilities to detect future ones, avoiding the unrealistic scenario of using future knowledge to predict past vulnerabilities.

\textbf{LLM-as-Judge reliability.} Following CORRECT, we employ LLM-as-Judge to evaluate reasoning correctness. We acknowledge that using LLMs to judge other LLMs' outputs may introduce bias. 
\begin{change}
To calibrate the judge's accuracy, two authors manually evaluated 50 randomly sampled cases and obtained 82\% agreement with the GPT-5 judge, consistent with the reliability reported in the original CORRECT paper (92\% with GPT-4o). A detailed analysis of the disagreement cases is provided in Appendix~G. Overall, we consider LLM-as-Judge an acceptable evaluation choice for reasoning correctness, as no ground-truth data-flow and control-flow annotations currently exist.
\end{change}


\section{Conclusion}\label{sec:conclusion}

In this work, we introduced VulInstruct, a specification-guided approach that systematically mines security specifications from historical vulnerabilities to enhance vulnerability detection. By combining specifications extracted from patches and CVE records with recurring attack patterns across projects, VulInstruct enables large language models to reason and align with implicit expectations of safe behavior. Our extensive evaluation on the PrimeVul benchmark demonstrates substantial improvements in F1-score, recall, and pair-wise discrimination compared with state-of-the-art baselines. More importantly, VulInstruct has shown practical utility by discovering a previously unknown high-severity vulnerability in real-world software. These results highlight the value of bridging implicit security expertise with automated analysis, opening new directions for integrating structured security knowledge into LLM-based vulnerability detection.

\bibliographystyle{ACM-Reference-Format}
\bibliography{my_paper/references}

\clearpage
\appendix

\section{CORRECT Evaluation Framework}

The CORRECT evaluation introduces rationale-based assessment to verify whether models identify vulnerabilities for the correct reasons:
The rationale refers to the model's step-by-step reasoning process and results that lead to its vulnerability detection conclusion. CORRECT evaluates this rationale against ground-truth vulnerability information (CVE descriptions, patches, and commit messages) to determine correctness:
The framework operates as follows:
\begin{enumerate}
    \item \textbf{For vulnerable code}: The model must correctly identify the ground-truth vulnerability in its rationale. If the rationale includes the actual vulnerability cause or key elements in vulnerabilities, it is deemed correct; otherwise, the detection is considered a false negative despite potentially correct binary labeling.
    \item \textbf{For patched code}: Two evaluation modes are employed:
    \begin{itemize}
        \item \textit{Lenient Mode}: Accepts the model's results if it either correctly identifies the code as non-vulnerable, or if it reports a vulnerability but the rationale does not reference the original (now patched) vulnerability. 
        \item \textit{Strict Mode}: This mode extends Lenient Mode by adding an explicit iterative check. When the model flags a patched function as vulnerable but does not reference the ground-truth vulnerability, the framework instructs the model to disregard the previously reported issues and re-evaluate whether the function still contains a vulnerability. If during this iterative process the model again reports the ground-truth vulnerability, the case is marked as a false positive, since the model fails to recognize that the vulnerability has already been fixed.
    \end{itemize}
\end{enumerate}
These scenarios are summarized in Table~\ref{tab:correct-modes}, which contrasts the evaluation outcomes of Lenient and Strict modes across different prediction cases.
In practice, this rationale assessment is implemented by another LLM acting as a judge, which takes the model’s explanation as input and decides whether it correctly reflects the ground-truth vulnerability.


\begin{table}[h]
\centering
\caption{Evaluation outcomes under Lenient and Strict modes for different prediction scenarios.}
\label{tab:correct-modes}
\begin{tabular}{llllcc}
\toprule
\textbf{Ground Truth} & \textbf{Prediction} & \textbf{Rationale} & \textbf{Lenient} & \textbf{Strict} \\
\midrule
Vulnerable & Vulnerable & Correct & TP & TP \\
Vulnerable & Vulnerable & Incorrect & FN & FN \\
Vulnerable & Non-vulnerable & -- & FN & FN \\
Patched & Non-vulnerable & -- & TN & TN \\
Patched & Vulnerable & References original vuln & FP & FP \\
Patched & Vulnerable & Other issues only & TN & Feedback$\rightarrow$TN/FP \\
\bottomrule
\end{tabular}
\end{table}

\textbf{Choice of Evaluation Mode.} While CORRECT provides both Lenient and Strict modes, we adopt Lenient Mode in our evaluation based on empirical evidence and practical considerations. 

Empirical results in the CORRECT paper, including extensive statistics reported in its appendix, show that the benefits of \textit{Strict Mode} are marginal: only 10.3\% of cases are modified after the first feedback round, dropping to 3.6\% in round 2, and merely 1.5\% by round 4. This diminishing return suggests that the iterative refinement process yields small improvements while significantly increasing computational costs.

In addition, applying \textit{Strict Mode} to existing methods is particularly problematic. Training-based approaches such as \textbf{ReVD}, and multi-agent frameworks such as \textbf{VulTrial}, rely on carefully designed instructions. 
Forcing these models to ignore previously reported vulnerabilities and repeatedly re-analyze the code disrupts their prompt structures, often leading to poor performance. 
Supporting such feedback loops would require retraining models from scratch with feedback-aware objectives, which introduces substantial overhead in both data construction and training optimization.

Given these issues, we adopt \textit{Lenient Mode} for all evaluations. This ensures a fair comparison across baselines while still keeping evaluation rigor through rationale correctness assessment.

\section{Prompt Template.}

\begin{promptbox}[Prompt: System Prompt for General Specifications]
A structured threat modeling analysis process where security experts conduct systematic security analysis based on provided information. The expert must:

\textbf{1. Understand Code Context} (within \texttt{<understand>} tags)

Thoroughly analyze and describe the system context without revealing the vulnerability itself:

\textbf{System Identification}
\begin{itemize}[leftmargin=*, noitemsep, topsep=0pt]
    \item \textbf{What system}: Clearly identify the software system, library, or application
    \item \textbf{Domain/Subsystem}: Specify the particular domain or subsystem where the code operates  
    \item \textbf{Module/Component}: Identify the specific module, component, or functional unit
\end{itemize}

\textbf{Functional Analysis}
\begin{itemize}[leftmargin=*, noitemsep, topsep=0pt]
    \item \textbf{Core functionality}: Describe what this system/module is designed to do in detail: 1. 2. 3.
\end{itemize}

\textbf{2. Security Domain Classification} (within \texttt{<classification>} tags)

Classify vulnerabilities according to 10 core security domains:

\textbf{Core Security Domains:}
\begin{enumerate}[leftmargin=*, noitemsep, topsep=0pt]
    \item \textbf{MEM}: Memory Safety [Buffer errors, pointer issues, use-after-free, allocation problems, etc.]
    \item \textbf{STATE}: State Management [Inconsistent states, object lifecycle, concurrency issues, etc.]
    \item \textbf{INPUT}: Input Validation [Parsing logic, data validation, type checking, encoding, etc.]
    \item \textbf{LOGIC}: Program Logic [Arithmetic errors, type confusion, logical mistakes, etc.]
    \item \textbf{SEC}: Security Features [Authentication, cryptography, permissions, policy enforcement]
    \item \textbf{IO}: I/O Interaction [Filesystem operations, networking, device interaction, etc.]
    \item \textbf{CONF}: Configuration Environment [Configuration parsing, environmental variables, etc.]
    \item \textbf{TIMING}: Timing \& Concurrency [Race conditions, synchronization issues, TOCTOU, etc.]
    \item \textbf{PROTOCOL}: Protocol Communication [Message parsing/formatting, session handling, etc.]
    \item \textbf{HARDWARE}: Hardware \& Low-level [Low-level interfaces, architectural specifics, etc.]
\end{enumerate}

\textbf{3. Security Specification} (within \texttt{<spec>} tags)

Security Specification helps understand how vulnerable code violates developer's original constraints and how patches implement fixes.

\hrulefill

\textbf{Example:}

\textbf{Input Information}
\begin{itemize}[leftmargin=*, noitemsep, topsep=0pt]
    \item \textbf{Repository}: ksmbd
    \item \textbf{Commit Message}: ksmbd: Fix dangling pointer in krb\_authenticate
    \item \textbf{CVE Description}: In the Linux kernel, the following vulnerability has been resolved: ksmbd: Fix dangling pointer in krb\_authenticate...
    \item \textbf{CWE Type}: CWE-416 (Use After Free)
\end{itemize}

\textbf{Code Diff:}
\begin{verbatim}
-if (sess->state == SMB2_SESSION_VALID)
+if (sess->state == SMB2_SESSION_VALID) {
     ksmbd_free_user(sess->user);
+    sess->user = NULL;
+}
\end{verbatim}

\textbf{Expected Output Format:}

\texttt{<understand>}\\
\textbf{System Identification}
\begin{itemize}[leftmargin=*, noitemsep, topsep=0pt]
    \item \textbf{What system}: ksmbd - in-kernel SMB server implementation for Linux
    \item \textbf{Domain/Subsystem}: SMB/CIFS network file sharing protocol implementation
    \item \textbf{Module/Component}: Kernel component receives SMB requests, uses netlink IPC...
\end{itemize}
\texttt{</understand>}

\texttt{<classification>}\\
\quad\texttt{<primary>MEM.LIFECYCLE</primary>}\\
\quad\texttt{<tags>[STATE.CONSISTENCY, SEC.AUTHENTICATION, PROTOCOL.SMB]</tags>}\\
\quad\texttt{<reasoning>The root cause is the failure to manage the lifecycle...</reasoning>}\\
\texttt{</classification>}

\texttt{<spec>HS-MEM-001: Pointer release operations require atomic cleanup with immediate nullification</spec>}\\
- Reasoning: Dangling pointer vulnerability → freed but not nullified → atomic release-nullification prevents use-after-free

\hrulefill

\textbf{Current Analysis Target:}

\textbf{Repository}: \{repository\}\\
\textbf{Commit Message}: \{commit\_message\}\\
\textbf{CVE Description}: \{cve\_description\}\\
\textbf{CWE Type}: \{cwe\_type\}\\
\textbf{Vulnerable Code}:
\begin{verbatim}
{vuln}
\end{verbatim}
\textbf{Solution}:
\begin{verbatim}
{fixed}
\end{verbatim}

Please conduct analysis following the above format.
\end{promptbox}

\begin{promptbox}[Prompt: Detailed Vulnerability Cases in General Specifications]
\footnotesize 
A structured threat modeling analysis process where security experts conduct systematic security analysis based on provided information.

\textbf{Analysis Framework}

\textbf{1. System Understanding} (provided context)\\
\texttt{\{understand\}}

\textbf{2. Security Specifications} (provided rules)\\
\texttt{\{specification\}}

\textbf{3. System-Level Threat Modeling} (within \texttt{<model>} tags)

Analyze vulnerability at system design level:
\begin{itemize}[leftmargin=*, noitemsep, topsep=2pt]
    \item \textbf{Trust Boundaries}: Identify where system components transition between trusted/untrusted states
    \item \textbf{Attack Surfaces}: Focus on realistic attack vectors that led to this specific vulnerability
    \item \textbf{CWE Analysis}: Trace complete vulnerability chain (e.g., initial CWE-X triggers subsequent CWE-Y, where at least one matches: \texttt{\{cwe\_type\}})
\end{itemize}

\textbf{4. Code-Level Analysis}

\textbf{Vulnerability Context} (within \texttt{<vuln>} tags)

Provide a granular, narrative explanation of the vulnerability:
\begin{enumerate}[leftmargin=*, noitemsep, topsep=2pt]
    \item \textbf{Entry Point \& Preconditions}: Describe how the attack is initiated and what system state is required
    \item \textbf{Vulnerable Code Path Analysis}: Step-by-step trace of execution flow, naming key functions and variables. Pinpoint \textbf{The Flaw} and its \textbf{Consequence}
    \item \textbf{Specification Violation Mapping}: Link code path steps to specific \texttt{HS-} specifications they violate
\end{enumerate}

\textbf{Fix Implementation} (within \texttt{<solution>} tags)

Explain how the patch enforces security specifications:
\begin{itemize}[leftmargin=*, noitemsep, topsep=2pt]
    \item Specific code changes and their security impact
    \item How fixes restore compliance with violated specifications
\end{itemize}

\hrulefill

\textbf{Example Output Format:}

\texttt{<model>}
\begin{itemize}[leftmargin=*, noitemsep, topsep=2pt]
    \item \textbf{trust\_boundaries}: User-Kernel boundary during SMB2 session setup; Intra-kernel function contract violation
    \item \textbf{attack\_surfaces}: Malicious SMB2 SESSION\_SETUP request; Error path exploitation
    \item \textbf{cwe\_analysis}: Primary CWE-416 (Use After Free) enabled by state management violation
\end{itemize}
\texttt{</model>}

\texttt{<vuln>}
\begin{enumerate}[leftmargin=*, noitemsep, topsep=2pt]
    \item \textbf{Entry Point}: Privileged user sends Netlink message with crafted CIPSOV4 tags
    \item \textbf{Code Path}: Loop processes tags → \textbf{The Flaw}: Off-by-one error in bounds check → \textbf{Consequence}: Stack buffer overflow
    \item \textbf{Violations}: \texttt{HS-MEM-001} (incorrect bounds check), \texttt{HS-STATE-002} (incomplete initialization)
\end{enumerate}
\texttt{</vuln>}

\texttt{<solution>}\\
\textbf{Change 1: Bounds Check Correction}
\begin{verbatim}
-if (iter > CIPSO_V4_TAG_MAXCNT)
+if (iter >= CIPSO_V4_TAG_MAXCNT)
\end{verbatim}
\textit{Compliance}: Changes exclusive to inclusive comparison, preventing array overflow

\textbf{Change 2: Complete Array Initialization}
\begin{verbatim}
-doi_def->tags[iter] = CIPSO_V4_TAG_INVALID;
+while (iter < CIPSO_V4_TAG_MAXCNT)
+    doi_def->tags[iter++] = CIPSO_V4_TAG_INVALID;
\end{verbatim}
\textit{Compliance}: Ensures all array elements initialized to safe values\\
\texttt{</solution>}

\hrulefill

\textbf{Input Information:}
\begin{itemize}[leftmargin=*, noitemsep, topsep=2pt]
    \item \textbf{CVE}: \texttt{\{cve\_description\}}
    \item \textbf{CWE}: \texttt{\{cwe\_type\}}
    \item \textbf{Commit}: \texttt{\{commit\_message\}}
    \item \textbf{Vulnerable Code}: \texttt{\{vuln\}}
    \item \textbf{Fixed Code}: \texttt{\{fixed\}}
    \item \textbf{Code Context}: \texttt{\{code\_context\}}
\end{itemize}

Please conduct analysis following the above framework.
\end{promptbox}

\begin{promptbox}[Prompt: VulInstruct Knowledge Scoring Mechanism]
You are a security expert. Please evaluate the relevance between the following code and VulInstruct vulnerability cases.

\textbf{Target Code}
\begin{verbatim}
{code_snippet}
\end{verbatim}

\textbf{VulInstruct Cases to Evaluate}\\
\texttt{\{chr(10).join(cases\_for\_evaluation)\}}

Please score the relevance of each case to the target code (1-10 points):

\textbf{Scoring Criteria:}
\begin{itemize}[leftmargin=*, noitemsep, topsep=2pt]
    \item \textbf{10 points}: Highly relevant, vulnerability type, trigger conditions, and code patterns are almost identical
    \item \textbf{8-9 points}: Strong relevance, main vulnerability features are similar, can provide valuable reference
    \item \textbf{6-7 points}: Moderate relevance, some features are similar, has certain reference value
    \item \textbf{4-5 points}: Weak relevance, only few similarities
    \item \textbf{1-3 points}: Very low relevance, basically no reference value
\end{itemize}

Please strictly follow the HTML format for output:

\begin{verbatim}
<vulinstruct_evaluation>
<case_1_score>6</case_1_score>
<case_1_reasoning>Scoring reason</case_1_reasoning>
<case_2_score>8</case_2_score>
<case_2_reasoning>Scoring reason</case_2_reasoning>
...
</vulinstruct_evaluation>
\end{verbatim}
\end{promptbox}


\begin{promptbox}[Prompt: Domain-specific Specification Extraction]
You are a security expert. Analyze these related vulnerabilities and extract reusable security specifications.

\textbf{Related Historical Vulnerabilities}\\
\texttt{\{chr(10).join(nvd\_descriptions)\}}

\textbf{Task: Extract Attack-Derived Specifications}

For each vulnerability pattern you identify:
\begin{enumerate}[leftmargin=*, noitemsep, topsep=2pt]
    \item \textbf{Identify the recurring attack mechanism} across these CVEs
    \item \textbf{Convert it to a positive security specification} that would prevent such attacks
    \item \textbf{Format as defensive requirements} developers must implement
\end{enumerate}

\textbf{Output Format:}
\begin{verbatim}
<attack_specifications>
  <specification_1>
    <attack_pattern>
      Description of recurring attack mechanism 
      in cve-xxx and cve-xxx in detail
    </attack_pattern>
    <defensive_spec>
      AS-DOMAIN-001: Security rule that describes 
      the code behavior that prevents this attack
    </defensive_spec>
    <implementation_hint>
      Specific checks or validations needed
    </implementation_hint>
  </specification_1>
  <specification_2>...</specification_2>
</attack_specifications>
\end{verbatim}
\end{promptbox}

\begin{promptbox}[Prompt: Vulnerability Detection]
You are a senior code security expert. Please perform systematic multi-layer security analysis on the following code.

\textbf{Analysis Mode:} \textit{[Determined by knowledge relevance scoring]}
\begin{itemize}[leftmargin=*, noitemsep, topsep=2pt]
    \item \textbf{Autonomous Analysis}: Low relevance with knowledge base, perform independent analysis
    \item \textbf{Knowledge-Assisted}: High relevance knowledge filtered through LLM evaluation as reference
\end{itemize}

\textbf{Input Components:}
\begin{itemize}[leftmargin=*, noitemsep, topsep=2pt]
    \item Code Snippet: \texttt{\{code\_snippet\}}
    \item Code Context: \texttt{\{code\_context\}}
    \item LLM-filtered Security Knowledge: \texttt{\{selected\_knowledge\}}
\end{itemize}

\textbf{Multi-Layer Vulnerability Analysis Framework}

\textbf{1. Surface Symptom Analysis}
Identify direct suspicious operations.

\textbf{2. Root Cause Investigation}
Trace deeper causes that give rise to the surface symptoms, focusing on data/control flow, completeness of input validation, adequacy of error handling, and potential attacker exploitation paths. 

\textbf{3. Architectural \& Contextual Analysis}
Examine broader design-level factors and domain-specific assumptions in the application logic.

\textbf{Comprehensive Security Assessment.}
Based on the above LLM-filtered Security Knowledge and three-layer analysis mode framework, please provide your professional judgment:

(i) Analysis Process: 
[Please describe your three-layer analysis process in detail, including discovered issues and reasoning chains]

(ii) Key Findings:
[List the most important security findings]

(iii) Final Conclusion: \\
Please strictly follow the format below for output:\\
\textbf{Output Format:}
\begin{verbatim}
<vulnerability_assessment>\\
Please strictly follow the format below for output:
  <has_vulnerability>yes/no</has_vulnerability>
  <confidence>0-1</confidence>
  <suspected_root_cause>Core findings summary</suspected_root_cause>
</vulnerability_assessment>
\end{verbatim}

\textbf{Format Description:}
\begin{itemize}[leftmargin=*, noitemsep, topsep=2pt]
    \item \texttt{has\_vulnerability}: ``yes'' or ``no''
    \item \texttt{confidence}: Confidence level between 0.0 and 1.0
    \item If a fixing solution has been applied, you may judge ``no''
    \item Focus on analysis quality, avoid over-sensitivity
\end{itemize}
\end{promptbox}

\section{Failure Analysis in Vul-RAG and ReVD}


We conduct a targeted analysis of the key performance factors in three representative approaches: the retrieval mechanism in Vul-RAG, the reasoning capability in ReVD, and the specifications in VulInstruct.

\textit{Failure Analysis of Vul-RAG.} Vul-RAG conducts vulnerability detection in an iterative retrieval manner. 
Given a target code snippet, the model first retrieves the top-$k$ most relevant vulnerability–patch cases ($k=10$ in our experiments). 
For each case, it checks (i) whether the same type of vulnerability exists in the target code 
and (ii) whether the corresponding patch has already been applied. 
The outcome is a binary signal: $(1,0)$ indicates a vulnerability without a patch (\emph{classified as vulnerable}), 
$(0,1)$ indicates a patch without vulnerability (\emph{classified as secure}), 
while $(0,0)$ or $(1,1)$ are inconclusive and trigger the next iteration. 
The process continues until a conclusive decision is made; if no decisive signal is found after all iterations, the output is treated as \emph{no decision}. As shown in Table~\ref{tab:baseline_failures}, over half of the samples (53.1\%) terminate within the first three iterations, while 12.7\% end without any decision.

\begin{table*}[t]
\centering
\caption{Failure analysis of baseline methods. Left: iteration distribution of Vul-RAG. Right: categorized failure cases of ReVD.}
\label{tab:baseline_failures}
\begin{minipage}{0.48\linewidth}
\centering
\begin{tabular}{lccc}
\toprule
\textbf{Iteration Range} & \textbf{Ratio} & \textbf{Vuln.} & \textbf{Secure} \\
\midrule
1--3   & 53.1\% & 12.5\% & 87.5\% \\
4--6   & 22.6\% & 18.0\% & 82.0\% \\
7--9   & 6.2\%  & 27.5\% & 72.5\% \\
10 (final) & 5.4\% & 50.0\% & 50.0\% \\
10 (no decision) & 12.7\% & 0.0\% & 100\% \\
\bottomrule
\end{tabular}
\caption*{(a) Vul-RAG iteration distribution where Vuln and Secure represents the final binary prediction.}
\end{minipage}\hfill
\begin{minipage}{0.48\linewidth}
\centering
\begin{tabular}{lc}
\toprule
\textbf{Failure Type} & \textbf{Ratio} \\
\midrule
Incorrect vulnerability categorization & 50\% \\
\quad Wrong CWE family & 30\% \\
\quad Similar type confusion & 20\% \\
Over-generalization & 26\% \\
Mechanism misinterpretation & 14\% \\
Others (incl.\ zero reasoning) & 10\% \\
\bottomrule
\end{tabular}
\caption*{(b) ReVD failure categories (50 manually analyzed cases).}
\end{minipage}
\end{table*}

As shown in Table~\ref{tab:baseline_failures}(a), over half of the samples (53.1\%) terminate within three iterations, while 12.7\% never reach a decision. A common failure is prematurely classifying samples as secure once a single patch match is observed, even if other vulnerabilities remain. This reliance on cross-project case matching proves unreliable: few vulnerabilities recur in exactly the same form, resulting in only 3.4\% reasoning correctness under CORRECT. 

\textit{Failure Analysis of ReVD.} 
We manually analyzed 50 vulnerable cases where ReVD produced the correct binary 
label but failed in reasoning. Six recurring error types emerged 
(Table~\ref{tab:baseline_failures}(b)): 
\textbf{(i) Zero reasoning} (4\%), where no explanation was provided; 
\textbf{(ii) Wrong vulnerability type} (30\%), where the model identifies an 
incorrect CWE family altogether; 
\textbf{(iii) Similar type confusion} (20\%), where the CWE category is in the 
right family but confused with a closely related type (e.g., CWE-125 vs.\ CWE-787); 
\textbf{(iv) Mechanism misinterpretation} (14\%), where the vulnerability type is 
correct but the exploitation path is directionally wrong---pointing to code 
constructs or attack vectors that do not realistically reach the flaw; 
\textbf{(v) Over-generalization} (26\%), where only vague statements like 
``potential memory safety issues'' replace specific root causes; 
and \textbf{(vi) Other errors} (6\%), including hallucinated function names or 
irrelevant code references. 
Although ReVD leverages large-scale reasoning data, its generated explanations 
often lack the precision and depth required for faithful root cause identification, 
highlighting the key limitation of reasoning-distillation approaches: they improve 
surface-level consistency but do not guarantee genuine understanding of vulnerability 
mechanisms.


\section{Manual analysis of successful cases in VulInstruct}

We analyze VulInstruct's specification-guided approach using DeepSeek-R1, examining how specifications enhance vulnerability detection. The method achieves 16.7\% improvement in detecting actual vulnerabilities (FN→TP) and 8.0\% reduction in false positives (FP→TN). To understand these improvements, we randomly sampled 10 cases from each category and performed a manual analysis, identifying four distinct repair mechanisms through which specifications enhance detection capabilities. Table~\ref{tab:repair_mechanisms} shows four recurring \emph{repair mechanisms}: \textbf{(i) Missing Security Dimension (20\% of cases)} which occurs when models lack awareness of entire vulnerabilities. For example, in CVE-2021-37848, the model only checked \texttt{strncmp} for logical errors. Our specification addressed this by mandating constant-time comparisons for security-sensitive operations. \textbf{(ii) Domain-Specific Blindness (30\% of cases)} occurs when models detect potential defects but underestimate their severity in specific contexts. For instance, in CVE-2022-24214, the model identified an integer overflow in DNS code but classified it as low risk, failing to recognize that DNS TXT records are attacker-controlled and high-risk vectors. Our domain specification AS-DOMAIN-1 provided this essential context, upgrading the assessment from "possible issue" to "high-severity vulnerability." \textbf{(iii) Deep Reasoning Enhancement (25\% of cases)} improves models' ability to connect isolated risks into complete analysis of vulnerability mechanism. We provide a detailed case study in Figure~ \ref{fig:case-study}. 
\textbf{(iv) Secure Pattern Validation (25\% of cases, exclusively FP→TN)} enables models to recognize secure implementations and avoid false positives. For example, in CVE-2022-21654, the model mistakenly flagged the use of \texttt{EVP\_sha256} as insecure. Our specification HS-CRYPTO-003 confirmed this as correct cryptographic practice, reinforcing that knowledge of secure patterns is as critical as vulnerability detection.

\begin{table}[h]
\centering
\caption{Repair mechanisms by which specifications improve vulnerability detection}
\label{tab:repair_mechanisms}
\resizebox{\columnwidth}{!}{%
\begin{tabular}{@{}lcccp{5cm}@{}}
\toprule
\textbf{Repair Mechanism} & \textbf{FN→TP} & \textbf{FP→TN} & \textbf{Total} & \textbf{Primary Knowledge Sources} \\
\midrule
Missing Security Dimension & 4 & 0 & 4 (20\%) & General + Domain-specific \\
Domain-Specific Blindness & 3 & 3 & 6 (30\%) & Domain-specific + Detailed cases \\
Deep Reasoning Enhancement & 3 & 2 & 5 (25\%) & Domain-specific + Detailed cases \\
Secure Pattern Validation & 0 & 5 & 5 (25\%) & General  + Detailed cases \\
\bottomrule
\end{tabular}
}
\end{table}

\section{Using VulInstruct Finding real-world vulnerability}


\textbf{Case Study: From CVE-2021-32056 to a New Access Control Bypass.} 
We provide a detailed case study to illustrate how our specification-guided auditing framework can facilitate the discovery of new real-world vulnerabilities. 
Starting from CVE-2021-32056 in the Cyrus IMAP Server (\texttt{cyrusimap/cyrus-imapd}), a randomly selected vulnerability case from the CORRECT dataset, we successfully identified and reported a previously unknown high-severity flaw in the same codebase. CVE-2021-32056 allowed authenticated users to bypass intended access restrictions on server annotations, leading to potential replication stalls and service disruptions. The root cause was an improperly scoped permission check in \texttt{imap/annotate.c}, where the \texttt{maywrite} check was nested inside a conditional block and skipped when the mailbox pointer was \texttt{NULL}. From this case, our framework distilled three reusable security specifications, including the specification HS-SEC-001 (annotation write operations must always enforce strict privilege-based access control).  

Guided by this specification, we constructed an auditing agent workflow that simulated the workflow of a manual security audit: starting from the known flaw, generalizing its pattern, and searching the codebase for other instances where security checks might be incorrectly scoped. 

Our workflow first prompted a LLM to generate candidate \texttt{git grep} commands that could reveal similar patterns elsewhere in the repository. We employed Gemini-2.5-Pro, whose relatively large context window allowed us to supply extracted code fragments without requiring a more elaborate dynamic windowing design. 
The generated commands reflected the model’s reasoning about how the extracted specification might be violated in other parts of the codebase. In particular, the model focused on core database operations such as \texttt{store} and \texttt{delete}, which are security-critical under HS-SEC-001, and combined them with contextual information from the repository’s application context (e.g., session management, mailbox operations). In doing so, the model hypothesized potential validation points where access control checks might be inconsistently applied. The next stage of our workflow was to feed the retrieved code snippets back into the model for analysis under the extracted specifications. This allowed the model to reason about whether the conditional checks in each match were relevant to security enforcement.
Among four generated queries produced, one was especially effective, as shown below:

\begin{lstlisting}[language=bash]
git grep -p --all-match \
  -e 'if (mailbox)' \
  -e 'cyrusdb_store\|cyrusdb_delete' \
  -- '*.c'
\end{lstlisting}

This query rediscovered the already patched \texttt{write\_entry} function in \texttt{annotate.c}, thereby validating the search strategy, while simultaneously surfacing additional candidate sites. Furthermore, the model dismiss some false positives, such as matches in \texttt{imap/tls.c}, where the conditionals guarded resource management logic rather than access control, and therefore did not violate HS-SEC-001. 
At the same time, the model highlighted \texttt{imap/mboxlist.c} as a high-risk location, noting that several conditional branches in this file surrounded critical database operations such as \texttt{cyrusdb\_store} and \texttt{cyrusdb\_delete}. To further investigate, we selected this candidate and applied our Automatic Context Extraction Tool to retrieve the surrounding function bodies together with a depth-3 call chain, ensuring that the model could reason about how access control checks were propagated across related functions. Given this enriched context, the model conducted a more detailed analysis and identified problematic control-flow paths. Guided by this assessment, we performed a closer inspection of the extracted functions---\texttt{mboxlist\_update}, \texttt{mboxlist\_update\_entry\_full}, and \texttt{mboxlist\_renamemailbox}---which ultimately led to the discovery of a severe privilege bypass. 
The most critical finding emerged in \texttt{mboxlist\_renamemailbox}, a function with complex multi-path control flow. Here, we discovered that a \texttt{goto} statement transferred execution directly to the database update section, bypassing any privilege checks:

\begin{lstlisting}[language=C]
if (mbentry->mbtype & MBTYPE_INTERMEDIATE) {
    // ... destination checks ...
    goto dbupdate;  // BYPASSES permission check!
}

myrights = cyrus_acl_myrights(auth_state, mbentry->acl);
if (!isadmin && !(myrights & ACL_DELETEMBOX)) {
    return IMAP_PERMISSION_DENIED;
}
\end{lstlisting}

As a consequence, an authenticated user without \texttt{ACL\_DELETEMBOX} rights could still rename intermediate mailboxes. This represents a direct violation of HS-SEC-001: security checks were present but misplaced, enabling a privilege bypass. Conceptually, this flaw mirrors CVE-2021-32056—the same fundamental security specification was violated—but here the error manifested through control-flow misordering (\texttt{goto}) rather than conditional scoping. 

The potential consequences were significant: unauthorized mailbox renaming could disrupt shared mailbox hierarchies, shift shared folders into private namespaces, and cause denial-of-service for legitimate users. We responsibly disclosed the issue to the Cyrus IMAP development team, who confirmed the vulnerability, reproduced it via integration tests, and patched it by relocating the permission check before the special-case handling logic. We ultimately assigned CVE-2025-56538 to this vulnerability, though the detailed information will not be made public during the review process.

\section{Automatic Context Extraction Tool}

We implement a unified Commit URL parser that normalizes repository and patch commit metadata across heterogeneous hosts. Using a small set of hand-written matching rules, the parser recognizes both modern and legacy interfaces (e.g., GitHub, GitLab, Bitbucket, as well as cgit/gitweb deployments such as kernel.org and GNU Savannah), and deterministically maps each commit URL to a canonical triple \{\texttt{repo\_name}, \texttt{commit\_hash}, \texttt{clone\_url}\}. For example, kernel.org cgit links are remapped to stable GitHub mirrors to enable uniform downstream handling.

Building on this, we resolve---when available via the host's API---the canonical parent repository for each commit, thereby obtaining the repositories for both the patched and vulnerable versions. Following CORRECT, we derive target functions from the change lines and then use \texttt{joern} and \texttt{cflow} to extract the four context types described in Section~4.1.1. For large repositories (e.g., Linux, TensorFlow), we further augment the default flow with lightweight subsystem/component identification and module-boundary detection, \emph{seeded by a small set of manually curated anchors} (Linux: \texttt{net}/\texttt{fs}/\texttt{kernel}; TensorFlow: \texttt{core}/\texttt{python}/\texttt{compiler}) and simple boundary rules (\texttt{find\_module\_root}, \texttt{is\_module\_boundary}). The analysis employs staged filtering (header-include analysis $\rightarrow$ call-graph edges $\rightarrow$ symbol-to-file mapping) and adaptive file selection (targets{+}module, callers/callees, include neighbors, paired \texttt{.h}/\texttt{.c}), together with conservative caps (up to 3{,}000 files; depth $\leq$ 3) and graceful timeouts/recovery. These heuristics keep CPG construction incremental and tractable while preventing uncontrolled expansion.

Our final preprocessed dataset provides substantially richer context than PrimeVul, which only supplies the full file in which a vulnerable function resides. In contrast, our dataset captures the four distinct types of context described in Section~4.1.1, thereby enabling more fine-grained program analysis and downstream tasks.

\section{LLM-as-Judge Disagreement Analysis}
\label{appendix:judge}

We analyzed the 9 disagreement cases (out of 50) between human evaluators and the GPT-5 judge and identified two main sources. First, some disagreements arise from ambiguous exploitation paths: when patches are applied at upstream locations without accompanying PoC or sanitizer reports, pinpointing the exact vulnerable branch is difficult even for human experts, and the judge occasionally accepts a related but imprecise reasoning chain. Second, coarse-grained or incorrect CWE annotations in the dataset introduce judge bias---high-level CWE mappings make partial matches easier to accept, and a small number of samples contain annotation errors (e.g., out-of-bounds write labeled as out-of-bounds read) that we confirmed via external advisories.

\section{Quality Analysis of the Specification Knowledge Base}
\label{appendix:spec_quality}

To address concerns about the consistency, diversity, and reusability of the extracted specification knowledge base, we conduct a detailed qualitative analysis on specifications extracted under CWE-20 (Improper Input Validation). We select CWE-20 because it is one of the most broadly defined and frequently occurring CWE categories, and therefore one where contradictions would be most likely to arise. In total, we analyze 60 specifications extracted from CWE-20-related patches.

\subsection{Consistency Analysis}

By design, our extraction prompt instructs the LLM to describe the \emph{expected safe behavior} that the patch restores, rather than to enumerate prohibited operations. Because different patches for the same CWE typically address distinct code contexts and attack surfaces, these positively-framed specifications naturally capture \emph{complementary facets} of the same vulnerability class. We argue that this formulation inherently has a lower probability of producing logical contradictions than negative-form rules would.

Through pairwise similarity matching across all 60 specifications, we identified only 1 specification pair with potential conflict, and found zero cases of direct logical contradiction. The identified case of apparent conflict is as follows:\textbf{HS-IO-003} (extracted from a QEMU CVE): requires that network device emulation must strictly validate all guest-controlled queue parameters, which could imply thorough and potentially costly checking. \textbf{HS-TIMING-001} (extracted from a radvd CVE): requires that network protocol implementations must avoid blocking delays during message processing, which could imply limiting the depth of validation.

If both specifications were naively applied to the same function, one might perceive a contradiction between validation thoroughness and processing responsiveness. However, this apparent conflict dissolves entirely once specifications are combined with the system-level code understanding that VulInstruct constructs at detection time (Section~4.2). The QEMU specification targets the virtual machine monitor's device emulation subsystem, where queue parameter validation consists of constant-time bounds checks (e.g., array index and integer overflow guards) that do not introduce blocking. The radvd specification targets a user-space IPv6 router advertisement daemon, where the actual vulnerability (CVE-2011-3605) was caused by calling \texttt{mdelay()} in the main processing loop---a fundamentally different concern.

We verified that across the entire PrimeVul test set, these two specifications were never simultaneously retrieved for any target function. Furthermore, even if conflicting specifications were co-retrieved in rare cases, the LLM-based knowledge scoring stage (Section~4.3) evaluates each candidate against the specific target function before inclusion in the detection prompt, and is well-positioned to filter out context-irrelevant specifications. In practice, cross-system conflicts are naturally resolved by the retrieval and scoring mechanism and do not affect detection.

\subsection{Diversity Analysis}

Although all 60 analyzed specifications originate from CWE-20 (Improper Input Validation), they generalize well beyond input checking into diverse security domains. The distribution across our predefined security domains is as follows:

\begin{itemize}[leftmargin=1.1em]
    \item \textbf{INPUT} (input validation: parsing logic, type checking, etc.): 30\% (18 specifications)
    \item \textbf{STATE} (state management: object lifecycle, concurrency, etc.): 18\% (11 specifications)
    \item \textbf{PROTOCOL} (protocol communication: message parsing, session handling, etc.): 23\% (14 specifications)
    \item \textbf{MEM} (memory safety: buffer errors, use-after-free, etc.) and \textbf{TIMING} (timing and concurrency: race conditions, TOCTOU, etc.): 29\% (17 specifications)
\end{itemize}

This distribution suggests that our extraction workflow captures the deeper security implications behind input validation failures---such as state corruption, protocol violations, and memory safety issues---rather than merely summarizing the surface-level patch descriptions.

\end{document}